\documentclass[fleqn,usenatbib]{mnras}

\usepackage{newtxtext,newtxmath}

\usepackage[T1]{fontenc}

\DeclareRobustCommand{\VAN}[3]{#2}
\let\VANthebibliography\thebibliography
\def\thebibliography{\DeclareRobustCommand{\VAN}[3]{##3}\VANthebibliography}

\usepackage{graphicx}	
\usepackage{amsmath}	
\usepackage{amssymb}	

\usepackage[dvipsnames, usenames]{xcolor}
\definecolor{linkcolor}{rgb}{0.0,0.3,0.5}
\definecolor{darkgreen}{rgb}{0.0, 0.5, 0.0}
\definecolor{darkcyan}{rgb}{0.0, 0.5,0.5}

\usepackage[normalem]{ulem}  

\title[Bulk viscosity in neutron star mergers]{Projecting the likely importance of weak-interaction-driven bulk viscosity in neutron star mergers}

\author[Elias R. Most et al.]{Elias R. Most,$^{1,2,3}$
\thanks{emost@princeton.edu} Steven P. Harris,$^{4}$ \thanks{harrissp@uw.edu}
Christopher Plumberg,$^{5}$ \thanks{plumberg@illinois.edu}
Mark G. Alford, $^{6}$\thanks{alford@physics.wustl.edu} Jorge Noronha,$^{5}$ \thanks{jn0508@illinois.edu}\newauthor
Jacquelyn Noronha-Hostler,$^{5}$ \thanks{jnorhos@illinois.edu}
Frans Pretorius,$^{2,7}$ \thanks{fpretori@princeton.edu} Helvi Witek,$^{5}$ \thanks{hwitek@illinois.edu} and Nicol\'as Yunes$^{5}$ \thanks{nyunes@illinois.edu}
\\
$^{1}${Princeton Center for Theoretical Science, Princeton University,
Princeton, NJ 08544, USA}\\
$^{2}${Princeton Gravity Initiative, Princeton University, Princeton, NJ
08544, USA}\\
$^{3}${School of Natural Sciences, Institute for Advanced Study, Princeton,
NJ 08540, USA}\\
$^{4}${Institute for Nuclear Theory, University of Washington, Seattle, WA 98195, USA}\\
$^{5}${Illinois Center for Advanced Studies of the Universe, Department of Physics, University of Illinois at Urbana-Champaign, Urbana, IL 61801, USA}\\
$^{6}${Physics Department, Washington University in St.~Louis, Saint Louis, MO 63130, USA}\\
$^{7}${Department of Physics, Princeton University, Princeton, NJ 08544, USA}\\
}

\date{Accepted XXX. Received YYY; in original form ZZZ}

\pubyear{2021}

\begin{document}
\label{firstpage}
\pagerange{\pageref{firstpage}--\pageref{lastpage}}
\maketitle
\begin{abstract}
In this work, we estimate how much bulk viscosity driven by Urca processes is likely to affect the gravitational wave signal of a neutron star coalescence.
{{In the late inspiral, we show that bulk viscosity affects the binding energy at fourth post-Newtonian (PN) order. 
Even though this effect is enhanced by the square of the gravitational compactness, the coefficient of bulk viscosity is likely too small to lead to observable effects in the waveform during the late inspiral, {when only considering the orbital motion itself}.}}
{{In the post-merger, however, the characteristic time-scales and spatial scales are different, potentially leading to the opposite conclusion.}} 
We post-process data from a state-of-the-art equal-mass binary neutron star merger simulation to estimate 
the effects of bulk viscosity (which was not included in the simulation itself).
In that scenario, we find that 
bulk viscosity can reach high values in regions of the merger. We compute several estimates of how much it might directly affect 
the global dynamics
of the considered merger scenario, and find that it could become significant. Even larger effects could arise in different merger scenarios or in simulations that include non-linear effects. This assessment is reinforced by a quantitative comparison with relativistic heavy-ion collisions where such effects have been explored extensively.

\end{abstract}




\section{Introduction}

Collider experiments \citep{Adamczewski-Musch:2019byl} and compact astrophysical objects, such as neutron stars \citep{Page:2006ud}, probe the most extreme states of matter in the universe. With densities $\rho>10^{14}\, \rm g/cm^3$ and temperatures ranging from eV (cold neutron stars \citep{Guillot:2019ugf}) to tens to hundreds of MeV (mergers and heavy ion collisions \citep{Dexheimer:2020zzs,Motornenko:2021dbq}).  Since the equilibrium properties of baryon dense matter cannot yet be determined by first-principle calculations \citep{Philipsen:2012nu}, relating them to macroscopic properties of neutron stars offers a unique opportunity for constraining them with astrophysical observations, (see \citet{Ozel:2016oaf} and \citet{Lattimer:2015nhk} for recent reviews).
Indeed, X-ray observations have been suggested as a promising way to directly infer the radius of neutron stars and, hence, probe the underlying equation of state (EoS) of dense matter at very low temperatures $\sim \mathcal{O}(\mathrm{keV})$ (see e.g. \citet{2009PhRvD..80j3003O,Ozel:2010fw,Steiner:2010fz,Nattila:2017wtj,Miller:2019cac,Riley:2019yda,Miller:2021qha,Riley:2021pdl}).
On the other hand, the gravitational wave detections of merging binary neutron stars \citep{TheLIGOScientific:2017qsa,Abbott:2020uma} and their electromagnetic counterparts \citep{GBM:2017lvd} have also been used recently to infer constraints on the dense matter EoS (e.g. \citet{Annala:2017llu,Margalit:2017dij,Bauswein:2017vtn,Rezzolla:2017aly,Ruiz:2017due,Most:2018hfd,Raithel:2018ncd,Abbott:2018exr,Shibata:2019ctb,Most:2020bba,Nathanail:2021tay}). 
This is based on inferring the deformability under gravitational tides from the detected gravitational wave signal \citep{Flanagan:2007ix,Read:2009yp} (see also the review \cite{Baiotti:2019sew}). Crucially, this inference is done assuming that the neutron stars in the inspiral are inviscid and cold.  

Different from the inspiral, the collision of two neutron stars can give rise to temperatures of $80\, \rm MeV$ or more.
Hence, the post-merger evolution does not only probe the cold EoS of nuclear matter but is fundamentally impacted by finite-temperature effects \citep{Kastaun:2016yaf,Hanauske:2016gia,Perego:2019adq,Endrizzi:2019trv}.
It has been shown that the gravitational wave frequency spectrum is largely dominated by the cold part of the EoS \citep{Bauswein:2011tp,Bernuzzi:2012ci,Takami:2014zpa}. In fact, it has been suggested that these frequencies are not only characteristic for a given EoS, but that it might be possible to use them to determine the cold EoS with a sufficiently large number of post-merger gravitational wave detections \citep{Bose:2017jvk}.

Neglecting finite-temperature effects in this inference of the EoS might then lead to systematic errors, {by only taking cold physics into account}. 
Some of these errors can be associated with the appearance of new degrees of freedom such as hyperons \citep{Sekiguchi:2011mc,Radice:2016rys} and quarks \citep{Most:2018eaw,Bauswein:2018bma,Most:2019onn,Blacker:2020nlq}, and have been shown to modify the gravitational wave signal. It has further been suggested that Urca processes lead to weak-interaction-driven bulk viscosity in the early post-merger phase \citep{Alford:2017rxf,Alford:2019qtm}, potentially being able to damp the gravitational emission on timescales of $<10\, \rm ms$.
Additionally, small scale turbulence produced in the merger and large magnetic fields might be able to affect the gravitational wave emission by providing a rapid form of angular momentum transport in the newly formed neutron star merger remnant, which has been investigated by effective shear viscous prescriptions (\citet{Shibata:2017xht,Radice:2017zta,Duez:2020lgq}, see also \citet{Chabanov:2021dee}).

\citet{Alford:2017rxf} argued that neutrino-driven thermal transport and shear dissipation are unlikely to affect the post-merger gravitational wave signal in the millisecond range (unless small-scale turbulent motion occurs).
However, bulk viscosity appears to have greater potential importance. 
In neutrino-transparent $npe$ matter bulk viscous damping of density oscillations arises from Urca re-equilibration of flavor; the resultant damping time has been estimated \citep{Alford:2019qtm} and found to be in the millisecond time range for matter in certain density and temperature ranges\footnote{In the neutrino-trapped regime, $T\gtrsim 5\,\text{MeV}$ \citep{Roberts:2016mwj,Alford:2018lhf}, flavor equilibration is faster and bulk viscosity seems to be a small effect \citep{Alford:2019kdw,Alford:2020lla}.}. This is fast enough to affect the evolution of the merger
and potentially leave an imprint in the corresponding gravitational waves. {{If so, bulk viscosity may provide}} 
a unique opportunity to extract, for the first time, information about out-of-equilibrium\footnote{The {$npe^-$} matter in a neutron star merger is locally in thermal equilibrium, but can be driven out of chemical equilibrium - see the discussion in Sec.~\ref{sec:bulk}.  Neutrinos, depending on their mean free path in a given location in the merger, may or may not be thermally equilibrated \citep{Endrizzi:2019trv}.} properties of the hot and ultradense matter formed in binary neutron mergers.

In this work we go beyond the initial analysis done in   \citet{Alford:2017rxf} and \citet{Alford:2019qtm}. 
In Sec. \ref{sec:bulk}, we start with a calculation of bulk viscosity for a range of phenomenologically plausible equations of state, and use it to estimate the relative importance of bulk viscosity in the evolution of a neutron star merger.
{Specifically, in Sec. \ref{sec:inspiral} we provide an estimate for the importance of bulk viscosity in the inspiral, finding negligible imprints on the gravitational waveform. }Different from this, in Sec. \ref{sec:post}, we {provide a realistic estimate} of the bulk viscous contribution to the pressure in the background of a state-of-the-art (though non-dissipative) neutron star merger simulation, {thereby establishing that bulk viscosity is non-negligible in the post-merger evolution phase}.
This allows us to
{gauge} in which stages and regions of the merger bulk viscosity can be expected to be dynamically important and whether bulk viscosity will have a significant influence on gravitational waves emitted during binary coalescence. {
The bulk viscosity is obtained from a  different EoS from the one used in the merger simulation. To address this inconsistency we} consider viscosities computed for three different EoSs that reasonably cover the allowed parameter space.

To test if the magnitude of the bulk viscosity will be potentially influential on neutron star merger dynamics, in Sec. \ref{sec:hic} we make direct comparisons to state-of-the-art relativistic viscous hydrodynamic calculations from heavy-ion collisions.  Viscous effects have been incorporated (with full back-reactions and with coupling terms between shear and bulk viscosity) for over a decade in the field of heavy-ion collisions \citep{Romatschke:2017ejr} and have been well-constrained by hundreds of experimental data \citep{Bernhard:2019bmu}. Even a small bulk viscosity can influence the final experimental observables in heavy-ion collisions \citep{Monnai:2009ad,Song:2009rh,Bozek:2009dw,Dusling:2011fd,Noronha-Hostler:2013gga, Ryu:2015vwa,Ryu:2017qzn}. 

{{Throughout the rest of this paper, we use the following conventions. For the most part, we employ geometric units in which $G = 1 = c$, although we re-instate units in some cases to make contact with experiment. We also employ the Einstein summation convention and label components of spacetime vectors with Greek indices. Other conventions are consistent with those of~\citet{Misner:1973prb}.}}

\section{Bulk viscosity in nuclear matter}
\label{sec:bulk}
Bulk viscosity models the resistance experienced by matter to compression/expansion, which leads to an out-of-equilibrium correction to the pressure of the fluid and, consequently, to an increase in entropy \citep{Rezzolla_Zanotti_book}. In the context of neutron star mergers, the minimal set of equations that describe the evolution of a relativistic bulk viscous fluid is defined by the conservation of baryon number, $\nabla_\mu J^\mu=0$, where the baryon current is $J^\mu = \rho u^\mu$ with $\rho= m_b n$ being the rest mass density for baryon mass $m_b$ and baryon number density $n$, and $u^\mu$ the fluid's 4-velocity (normalized such that $u_\mu u^\mu=-1$). The energy-momentum tensor 
\begin{align}
T_{\mu\nu} = \left(e+P+\Pi\right) u_\mu u_\nu + (P+\Pi) g_{\mu\nu}
\end{align}
is conserved $\nabla_\mu T^{\mu\nu}=0$, where $e$ is the comoving energy density, $P$ is the equilibrium pressure defined by the equation of state $P=P(e,\rho)$, $g_{\mu\nu}$ is the spacetime metric, and $\Pi$ is the bulk scalar, which describes the out-of-equilibrium correction to the pressure due to bulk viscosity ($\Pi$ vanishes in equilibrium).
The conservation laws have to be solved together with Einstein's equations, so the dynamical variables of the system are (schematically) $e, \rho, u_\mu, \Pi$, and $g_{\mu\nu}$. In order to close the system of equations, one has to specify how $\Pi$ is dynamically obtained.   

Near equilibrium, $\Pi$ can be expanded in powers of the derivatives of the hydrodynamic variables. To first-order in derivatives, a proper account of the dynamics can be done using the generalized 1st-order theory proposed in \citet{Bemfica:2017wps,Bemfica:2019knx,Bemfica:2020zjp,Kovtun:2019hdm,Hoult:2020eho} (BDNK), which leads to a strongly hyperbolic set of equations of motion for the fluid coupled to Einstein's equations \citep{Bemfica:2020zjp}. Alternatively, when deviations from equilibrium are not small one may employ a 2nd-order approach, such as Israel-Stewart theory \citep{Israel:1979wp}, which was proven to be strongly hyperbolic in \citet{Bemfica:2019cop} when only bulk viscous effects are taken into account. 
{
In this paper, we only consider the leading order corrections coming from $\Pi$ in the relativistic Navier-Stokes limit \citep{Rezzolla_Zanotti_book} where
\begin{align}
\Pi \simeq -\zeta \nabla_\mu u^\mu,
\label{eqn:bulk_NS}
\end{align}
and $\zeta=\zeta(e,\rho)$ is the (dynamic) bulk viscosity transport coefficient.
It is useful to introduce a dimensionless quantity to characterize the properties of a viscous flow. This is typically done by means of the inverse Reynolds number, which is generally defined as
\begin{align}
    {\rm Re}^{-1} = \frac{{\bar \mu}/\bar\rho}{\bar{v} L},
    \label{eqn:Re}
\end{align}
where $\bar{v}$, $\bar{\rho}$, and $L$ are the characteristic velocity, density, and length scale of the underlying flow, and $\bar{\mu}$ is the associated dynamical viscosity, which is $\zeta$ for bulk viscous flows.} 
\\
We also introduce the (bulk) viscous ratio,
\begin{align}
    \chi = \frac{\Pi}{e+P},
    \label{eqn:bulk-viscous-ratio}
\end{align}
which in the non-relativistic limit reduces to $\chi\rightarrow \Pi/\rho$.
This quantity measures the local {\em direct} impact of bulk viscosity on the fluid variables, and it can be considered as a proxy for the inverse Reynolds number of the flow in the relativistic regime  \citep{Denicol:2012cn,Denicol:2019lio}. Taking the non-relativistic Navier-Stokes limit of \eqref{eqn:bulk-viscous-ratio}, and assuming that $\nabla_i \bar{v}^i \to 3L \bar{v}$, we can approximate the bulk viscous ratio as
\begin{align}
     \chi \approx 3 \,\bar{v}^2\, {\rm Re}^{-1}\,,
     \label{eqn:chi_ref}
\end{align}
where above $\rm Re^{-1}$ was defined using the microphysical viscosity $\zeta$. 

\subsection{Weak-interaction driven bulk viscosity}
On the millisecond timescales that are relevant to mergers, the bulk viscosity $\zeta$ of nuclear matter arises from the re-equilibration of the proton fraction (``beta equilibration'') via Urca (weak) interactions.  On the millisecond timescale, strong interactions ensure that the neutrons, protons, and electrons are always in thermal equilibrium, described by Fermi-Dirac momentum distributions\footnote{{There is an additional contribution to the bulk viscosity arising from nucleon scattering via the strong interaction to restore a thermal (Fermi-Dirac) distribution \citep{Kolomeitsev:2014gfa,SYKES19701}.  This contribution to the bulk viscosity} {is relevant on strong interaction timescales $\sim 10^{-23} \text{ s}$, but is negligible on the millisecond (and longer) timescale of neutron stars and mergers.}}. The departure from beta equilibrium is therefore captured in a single number, the isospin chemical potential $\mu_\Delta$ that relaxes to zero on a timescale $\tau$. 
In Fourier space, the bulk viscosity experienced by a low-amplitude density oscillation of angular frequency $\omega$ is
\begin{equation}
    \zeta(\omega) = A \dfrac{\tau}{1 + \omega^2\tau^2} \ ,
    \qquad \tau^{-1} = B \dfrac{\partial \Gamma_{n\to p}}{\partial \mu_\Delta} \Biggr|_{\parbox{2em}{\footnotesize beta\\[-0.5ex] equil}} \ .
\end{equation}
The prefactors $A$ and $B$ contain combinations of nuclear susceptibilities, which are determined by the equation of state
\citep{Alford:2019qtm,Alford:2017rxf,Alford:2010gw,Sawyer:1989dp}, and the net rate per unit volume of neutron to proton conversion $\Gamma_{n\to p}$ is obtained from a weak interaction rate calculation, which depends on the dispersion relations of the in-medium  neutron and proton excitations \citep{Alford:2019qtm,Alford:2018lhf,Yakovlev:2000jp}.
The bulk viscosity can be completely characterized by specifying the zero-frequency bulk viscosity $\zeta(0)=A\tau$ and the beta equilibration timescale $\tau$ as functions of density and temperature for matter in beta equilibrium.  From this, the frequency-dependent bulk viscosity is uniquely determined $\zeta(\omega)=\zeta(\omega\!=\!0)/(1+\omega^2\tau^2)$ \citep{Gavassino:2020kwo}.

\begin{figure}
    \centering
    \includegraphics[width=0.48\textwidth]{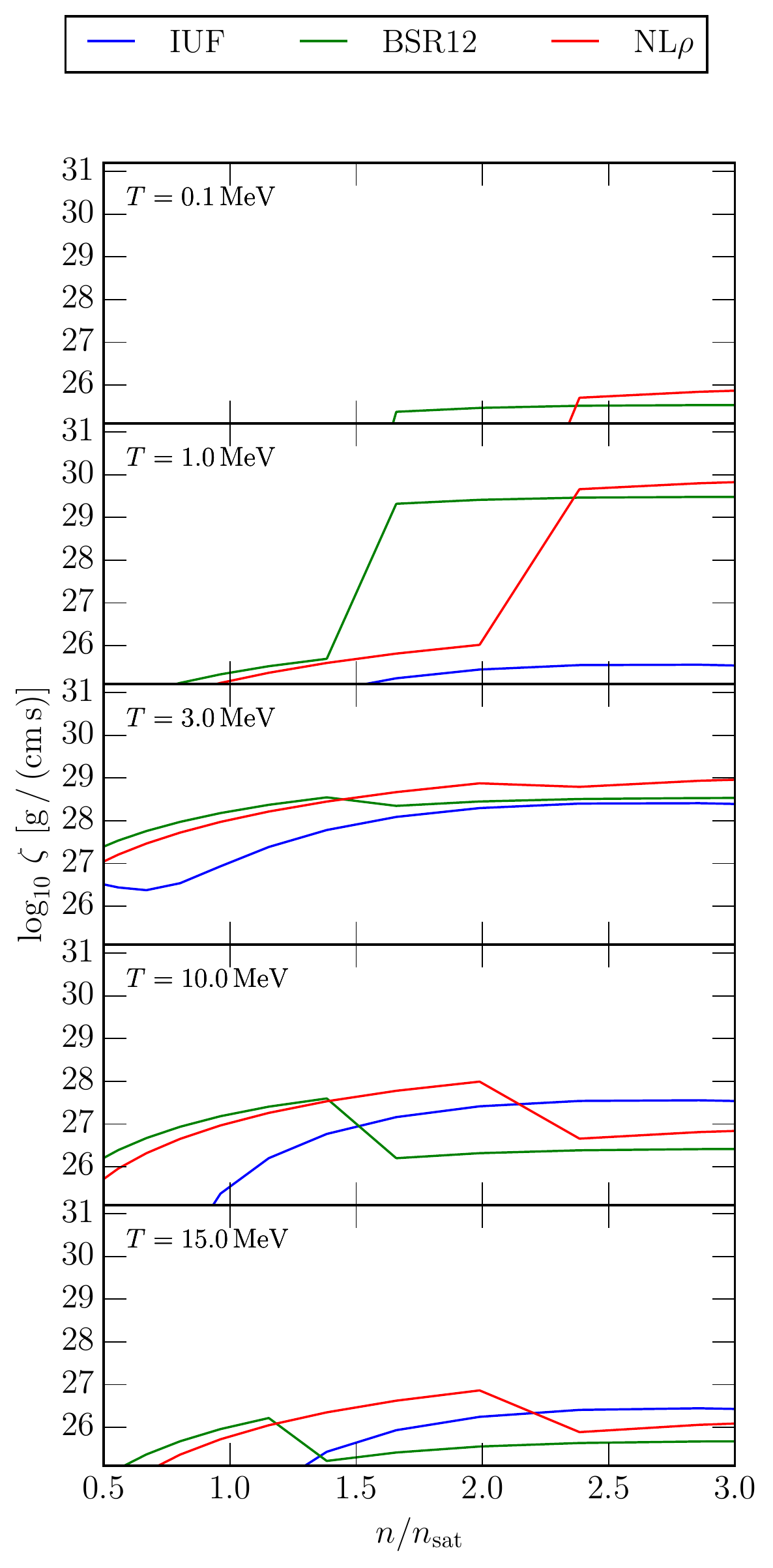}
    \caption{
     Bulk viscosity $\zeta$ for three different nuclear equations of state, for a small amplitude density oscillation $n(t) = n + \Delta n \cos{(\omega t)}$ where $\omega = 2\pi\times 1 \text{ kHz}$. The individual panels show slices for different temperatures $T$ and number densities $n$ in units of nuclear saturation density $n_{\rm sat}$.  Two of the equations of state have a direct Urca threshold in the displayed density range, causing dramatic changes in the bulk viscosity as a function of density.  Further explanation is given in the main text.}
    \label{fig:zeta}
\end{figure}

Calculations of the bulk viscosity and the energy dissipation it leads to are typically performed either for cold neutron stars, where beta equilibration is slow ($\tau^{-1} \ll \omega$) \citep{1968ApJ...153..835F,Sawyer:1989dp,1990ApJ...363..603C,Haensel:1992zz,Gupta:1997ce,Haensel:2000vz,Haensel:2001mw,Kolomeitsev:2014gfa}, or for hot protoneutron stars, where beta equilibration is fast ($\tau^{-1}\gg\omega$) \citep{Sawyer:1980wp} (see also the review article \cite{Schmitt:2017efp}).  It was only fairly recently that the thermodynamic conditions where bulk viscosity reaches its resonant maximum were sketched \citep{Lai:2001jt,Alford:2010gw,Alford:2019kdw}, and mapped out in great detail \citep{Alford:2019qtm,Alford:2020lla}.  One of the goals of this work is to establish whether neutron star mergers probe these thermodynamic conditions.  

{In the limit of low temperature where nuclear matter becomes strongly degenerate, the direct Urca processes become kinematically forbidden unless $p_{Fn}<p_{Fp}+p_{Fe}$.  This condition on the Fermi momenta becomes satisfied above the direct Urca threshold density.  Below the threshold density, beta equilibration occurs via the modified Urca processes $N+n\rightarrow N+p+e^-+\bar\nu_e$ and $N+e^-+p\rightarrow N+n+\nu_e$ ($N$ is either a neutron or a proton), which are much slower than direct Urca.  In strongly degenerate nuclear matter, the beta equilibration rate rises sharply when the density rises above the direct Urca threshold density.  The Urca rates in strongly degenerate nuclear matter are calculated in the ``Fermi surface approximation'' in \cite{Yakovlev:2000jp} (see also \cite{Alford:2018lhf}).}  In \citet{Alford:2019qtm}, the beta equilibration rate $\tau^{-1}$ was calculated by doing the full integration over phase space for the direct Urca process, which properly accounts for the gradual opening up of phase space at densities just below the direct Urca threshold \citep{Alford:2018lhf}.   

{For simplicity, in this work we use the Urca rates calculated in the Fermi surface approximation.  To mimic the blurring of the direct Urca threshold that occurs at finite temperature \citep{Alford:2018lhf}, we interpolate between the modified and direct Urca rates over a fixed density range around the direct Urca threshold density.  We have chosen this blurring of the threshold to be temperature-independent, but in reality the blurring increases with temperature \citep{Alford:2018lhf}.  {Additionally, although much of the matter in the merger remnant is at sufficiently high temperatures to trap neutrinos, for simplicity we continue to use the neutrino-transparent Urca rates.  This leads to an overestimate \citep{Alford:2019kdw,Alford:2020lla} of the bulk viscosity in the regions of the merger with temperatures above 5-10 MeV, but the bulk viscosity in these regions is insignificant (as will be seen in Figs.~\ref{fig:temp} and \ref{fig:chi}).}}

Due to current uncertainties in the nuclear equation of state, the weak
interaction rate (most importantly, the direct Urca threshold density
\citep{Reed:2021nqk,Beloin:2018fyp,Brown:2017gxd,Beznogov:2015ewa}) and the
nuclear susceptibilities are not well constrained.  To account for this, we
choose three nuclear equations of state, IUF \citep{Fattoyev:2010mx}, BSR12
\citep{Dhiman:2007ck}, and NL$\rho$ \citep{Liu:2001iz} with which to
calculate the bulk viscosity.  All three EoSs are based on relativistic
mean field theories \citep{Dutra:2014qga,Glendenning:1997wn}, where the
nucleons interact by exchanging mesons, whose fields are frozen to their
vacuum expectation values.  The three models differ in the types of meson
self-interactions included, as well as the values of the couplings, which
are obtained by fitting to properties of bulk nuclear matter or finite
nuclei.  The direct Urca threshold densities for IUF, BSR12, and NL$\rho$
are $4.1n_{\rm sat}$, $1.5n_{\rm sat}$, and $2.2n_{\rm sat}$ respectively.
In Fig.~\ref{fig:zeta}, we plot the bulk viscosity as a function of density
for these three EoSs.  Each panel shows a different temperature.  At low
temperatures, such as $T=1\text{ MeV}$, the beta equilibration rates are
slow compared to a 1 kHz density oscillation.  Below the direct Urca
threshold, the nuclear matter equilibrates through the modified Urca
process.  When the density rises above the direct Urca threshold, direct
Urca is now the process which drives the beta equilibration, and the beta
equilibration rate is now closer to (but still slower than) 1 kHz, causing
the bulk viscosity to rise dramatically above the direct Urca threshold.
The direct Urca threshold for IUF is at a higher density than we examine
here, so only modified Urca operates and there is no jump in the bulk
viscosity in the displayed density range.  For a higher temperature such as
$T=10\text{ MeV}$, the beta equilibration rate is fast compared to 1 kHz.
Below the threshold density, the modified Urca process equilibrates the
system quickly compared to the 1 kHz density oscillation.  When the density
rises above the direct Urca threshold, the beta equilibration rate becomes
even faster, causing a dramatic decrease in the bulk viscosity.

\section{Bulk viscosity in the inspiral}
\label{sec:inspiral}

\begin{figure*}
    \centering
    \includegraphics[width=\textwidth]{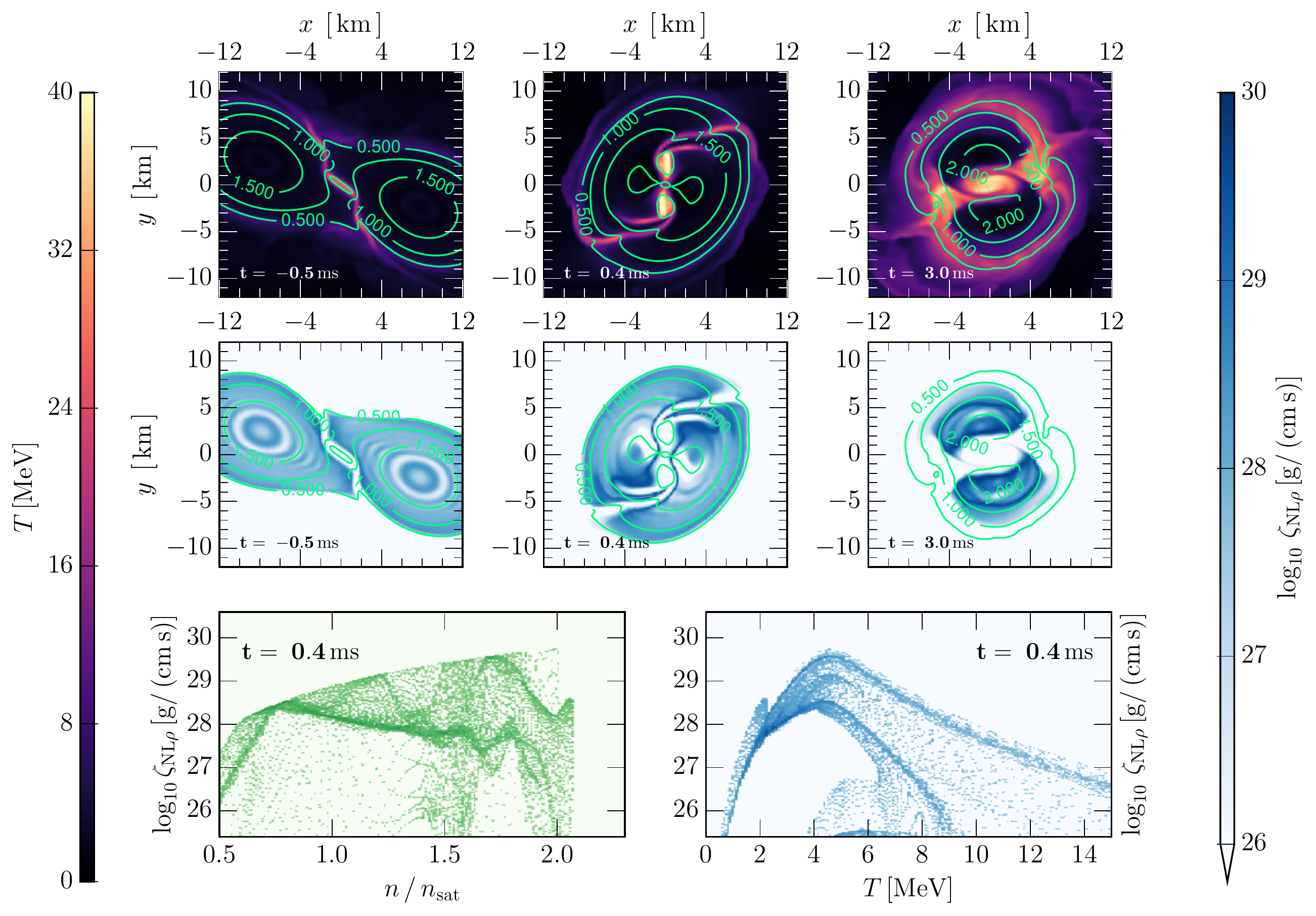}
    \caption{Thermodynamic conditions probed during the merger.
    {\it (Top row)} Three representative times during merger showing the temperature $T$ probed in the equatorial plane of the collision. The time $t$ is defined with respect to the time of merger.    The green lines are contours of baryon number density $n$ in units of nuclear saturation density $n_{\rm sat}$.
    {\it (Center row)} Same as above but showing the spatial distribution of bulk viscosities $\zeta_{\rm NL\rho}$ for the $\rm NL\rho$ model probed in the equatorial plane of the merger.
    {\it (Bottom row)} Distribution of bulk viscosities $\zeta_{\rm NL\rho}$ for the $\rm NL\rho$ model probed by the fluid elements during the merger in terms of baryon number density $n$ (left) and temperature $T$ (right) at $t=0.4$ ms, corresponding to the middle panels in the top and center rows.
    }
    \label{fig:temp}
\end{figure*}

Let us now estimate the impact of viscosity on the gravitational wave signal from the inspiral phase of a neutron star merger.  For an equal mass binary with neutron stars of mass $1.4 M_\odot$, the inspiral becomes observable to current detectors at orbital separations of about $700$ km, corresponding to a gravitational wave frequency of about $10$ Hz, as one can easily calculate from post-Newtonian (PN) theory~\citep{Blanchet:2013haa}. 
The inspiral phase ends just before the stars touch, which then corresponds to orbital separations of about $22$ km for stars with radii of $11$ km. During this phase, which lasts about 20 minutes, tidally induced fluid motions will heat up the stars.  Large-amplitude ``suprathermal'' bulk-viscous tidal heating originating from nucleonic Urca processes is expected to heat the stars up to, at most, tens of keV \citep{Arras:2018fxj}.  This is one order of magnitude higher than a previous estimate which included only shear viscosity \citep{Lai:1993di}, and is averaged over the star, so locally the temperature could be higher.  Hyperonic bulk viscosity is expected to be large at these temperatures, and could cause further dissipation \citep{Alford:2020pld} that could also lead to a continued increase in the average temperature of the star.  Finally, neutron star merger simulations see significant heating of the neutron star interiors during the last few orbits before merger, when the tidal deformation becomes large, finding the temperature to reach up to a few MeV (see \citep{Perego:2019adq} and Fig.~\ref{fig:temp}).  {At these MeV temperatures encountered in late inspiral,} the bulk viscosity in nuclear matter is expected to reach $\zeta \simeq (10^{27}$--$10^{29})\, \rm {g\,/\left( cm\,s\right)}$, as can be seen in Fig.~\ref{fig:zeta}.  

We now provide a first estimate of the impact of bulk viscosity on the dynamics of the inspiral. In doing so, we only consider the direct impact of the orbital motion on the star, but neglect secondary effects, such as oscillations of the neutron star \citep{Kokkotas:1999bd} and the back-reaction of viscosity that leads to tidal heating \citep{Arras:2018fxj,Lai:1993di}.  The {inspiraling} orbital motion induces compressional 
fluid motions, with fluid velocity $\left|v^i\right| = v^\phi \sim r \Omega$, where $r$ is the distance of the fluid element from the origin, {$\Omega = (m/r^3)^{1/2} \sim \left(v^\phi\right)^3$} is its orbital angular velocity, and $m$ is the total mass of the binary. { This directly implies that $v^\phi \sim r^{-1/2}$. In the following, we will drop the superscript and refer to $v^\phi$ as $v$. Additionally, the inspiraling motion of the star induces a radial velocity $v^r \sim r^{-3} \sim v^6$ \citep{Peters:1964zz}.}  Hence, the fluid velocity is therefore not the same everywhere inside the star, and one can approximate its spatial gradient via $|\nabla_\nu u^\mu| \sim v^i/{\cal{R}}$, where ${\cal R}$ is the scale of spatial variation.  Using ~\eqref{eqn:bulk_NS}, we can then estimate the impact of the bulk scalar in the inspiral via 
\begin{align}
    \left|\Pi\right| = \zeta \left|\nabla_\mu u^\mu\right| \approx \zeta \partial_r v^r \sim \zeta \, \left(\frac{v^r}{{\cal{R}}}\right) \sim \zeta v^5 \Omega \sim \zeta v^8,
    \label{eqn:Pi_insp}
\end{align}
where we have used that $v^i = u^i/u^0 \simeq u^i$ in the post-Newtonian approximation when $\left|v\right|\ll 1$. We have here also used that the scale of the derivatives is the orbital scale, so we can approximate ${\cal{R}} \sim r_{12}$, where {$r_{12}\sim r$} is the orbital separation; clearly, here we have used Kepler's third law and the virial theorem to write $v^\phi/{\cal{R}} \sim v/r_{12} \sim \Omega$. 

The impact of this bulk scalar on the inspiral phase can be estimated by comparing its contribution to the energy-momentum tensor. Just as one can compare the pressure to the energy density, $P/e$, to determine that pressure is a 1PN relative order modification to the inspiral, one can similarly compute the ratios defined by $\chi$ 
in ~\eqref{eqn:chi_ref} 
to estimate the impact of $\zeta$ 
in the inspiral. Using Kepler's Third law, we can relate the orbital separation and the velocity.  With this in hand, and approximating the mean density of a neutron star of mass $M$ and radius $R$ with $\rho \sim M/R^3$, we then find
\begin{align}
    \chi &\simeq \left(\frac{G}{c^3}\zeta R\right) \left(\frac{G\,M}{c^2 R}\right)^{-2} \left(\frac{v}{c}\right)^{8}\,, 
\nonumber \\
&= 
    7 \times 10^{-4} \left(\frac{\zeta}{10^{28} \; \rm{g} \; \rm{cm}^{-1} \; \rm{s}^{-1}}\right) \left(\frac{R}{11 \; \rm{km}}\right)  \left(\frac{0.19}{\mathcal{C}}\right)^2 \left(\frac{v}{c}\right)^{8} 
    \,,
    \label{eqn:Pirho}
\end{align}
where we have restored the factors of $G$ and $c$ here for clarity. Hence, we conclude that bulk 
viscous effects driven by the orbital motion should constitute a correction at 
4PN order,
which is suppressed by the dimensionless number $G \zeta R/c^3$ and $G \eta R/c^3$, while enhanced by two powers of the neutron star compactness $\mathcal{C} = G M/(c^2 R)$.

In this estimate we have assumed that $\zeta$
remains constant. As discussed above, tidal heating can increase the temperature of the stars during the inspiral, even leading to different mechanisms providing bulk viscosity at different times before merger. This might introduce an effective dependence of the bulk viscous stress on orbital velocity that will complicate the above order-of-magnitude PN analysis, and should be explored in more detail in future works. Similarly, oscillations of the star induced by the orbital motion might introduce higher-order velocity corrections $\left(v/\mathcal{R}\right)^n$ in ~\eqref{eqn:Pi_insp} 
, which would add additional higher-order PN terms to our estimate in ~\eqref{eqn:Pirho}. 
 
What impact will such an effect have on the gravitational waves emitted? The rate of change of the gravitational wave frequency is controlled by the binary's total energy $E$ and the rate of change of this energy via $df/dt = (dE/df)^{-1} (dE/dt)$. The rate of change of the energy is usually obtained from a balance law, $dE/dt = -{\cal{L}_{\rm GW}}  -{\cal{L}_{\rm diss}}$, where ${\cal{L}_{\rm GW}}$ is the gravitational wave luminosity and ${\cal{L}_{\rm diss}}$ the additional viscous energy dissipation. The effects discussed above would also impact the total energy, since the stress-energy tensor of the neutron stars would be modified by terms that depend on the bulk 
viscosity. It is not possible to know precisely what this modification will be without a detailed calculation, but we can again do a Fermi estimate. 
\begin{figure*}
    \centering
    \includegraphics[width=0.75\textwidth]{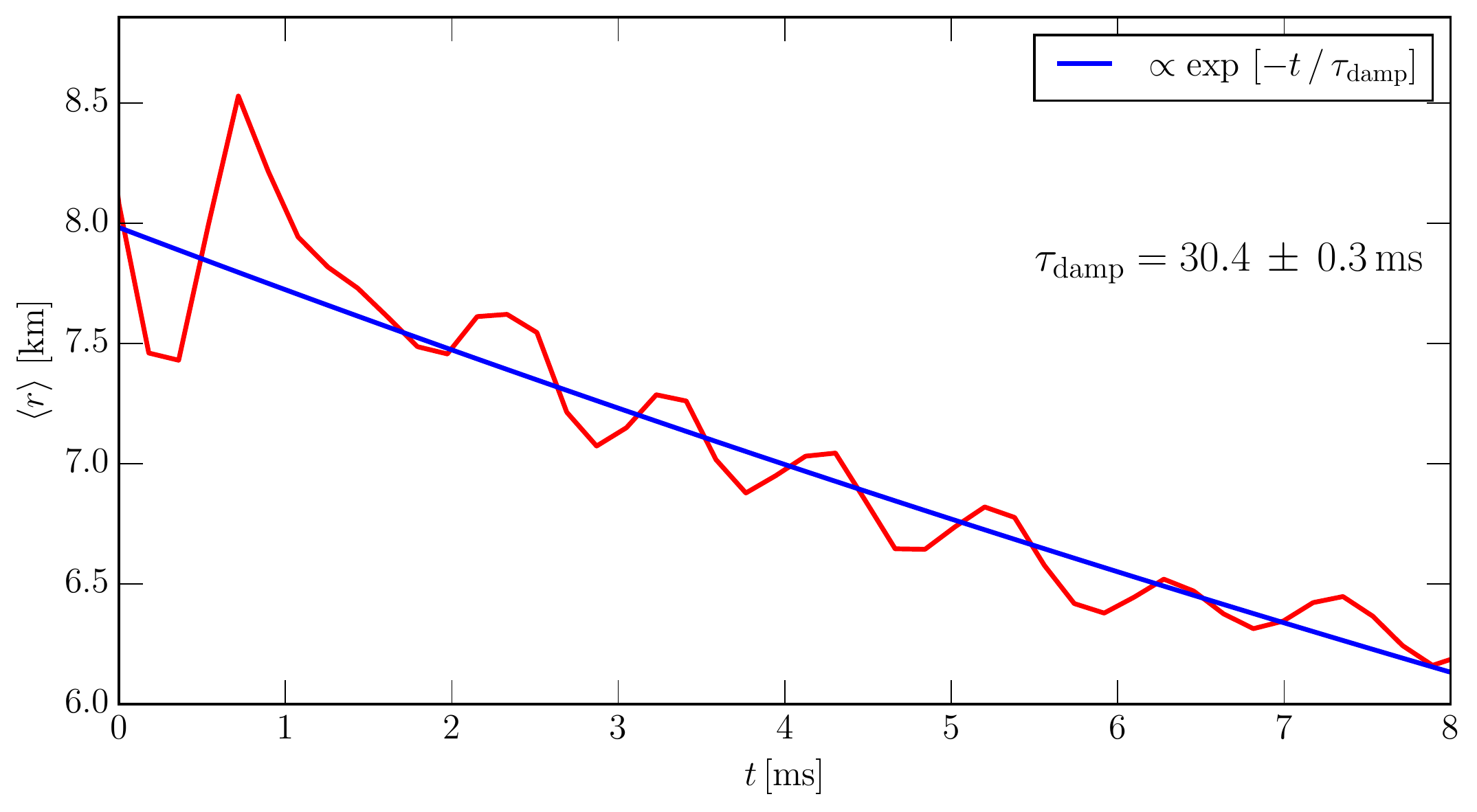}
    \caption{Radial position $\left<r\right>$ of the center of mass of one of the merging stellar cores.
    Due to the equal-mass nature of the binary $\left<r\right>$ is the same for both cores. The time $t$ is stated relative to the time of merger.
    Continued emission of gravitational waves will eventually dampen the oscillations which happen at a frequency of $\omega \simeq 1\, \rm kHz$. This damping occurs on a characteristic time scale $\rm \tau_{\rm damp}$. }
    \label{fig:pos}
\end{figure*}
Let us then focus on how bulk 
viscosity affect the total energy to estimate its impact on the gravitational wave phase. Using the fact that the energy-momentum tensor is corrected at 4PN order, we would expect a modification of the same PN order in the rate of change of the gravitational wave frequency and,
thus, on the gravitational wave phase. Assuming then that the gravitational wave phase acquires a term proportional to $\chi$ and i.e.~$\Psi \to \Psi [1 + {\cal{O}}(1/c^2) + \chi]$, we can then estimate the signal-to-noise ratio (SNR) that would be required to measure $\zeta$. Ignoring covariances for this Fermi estimate, we have that $1/{\rm{SNR}} = \Psi_{\rm{Newt}} (\Delta \zeta) \partial_\zeta \chi$, where $\Psi_{\rm{Newt}} = (3/128) (\pi {\cal{M}} f)^{-5/3}$ is the Newtonian part of the Fourier phase, with ${\cal{M}}$ the chirp mass, which then means that we could 
measure $\zeta$ to roughly 
\begin{align}
 \frac{\Delta \zeta}{\zeta} &\sim \left(\frac{1}{{\rm{SNR}}}\right)
 \left(\frac{1}{\Psi_{\rm{Newt}}}\right)
 \chi ^{-1} 
 \nonumber \\
 &= \left(\frac{1}{{\rm{SNR}}}\right) 
  \left(\frac{128}{3 (\pi {\cal{M}} f)^{-5/3}}\right)
 \left[\left(\zeta R\right) \left(\frac{M}{R}\right)^{-2} \left(\frac{v}{c}\right)^{8} \right]^{-1}\,,
\end{align}
Unfortunately, this shows the effect is incredibly small and essentially un-measurable in the inspiral {{unless the viscosity coefficients are significantly large}}. Let us evaluate the above estimate at a chirp mass corresponding to two equal mass neutron stars of $1.4 M_\odot$ and radius of $11$ km, at a gravitational wave frequency of 400 Hz (corresponding to a separation of about $60$ km). Then, {{if the bulk viscosity was only as large as  $\zeta = 10^{30} \; {\rm{g}} \; {\rm{cm}}^{-1} \; {\rm{s}}^{-1}$, then one would require an SNR of $10^5$ to get a $10\%$ measurement.}}  We see then even though there is an enhancement of the effect by two powers of the compactness, bulk viscosities {{of about $10^{30} \; {\rm{g}} \; {\rm{cm}}^{-1} \; {\rm{s}}^{-1}$ would not be}} measurable because {{their} effect} enters at too high a post-Newtonian order. 
{{We can, however, flip the argument around. Given an SNR of $10^2$, one should be able to measure or place an upper limit on $\zeta \lesssim 10^{33}  \; {\rm{g}} \; {\rm{cm}}^{-1} \; {\rm{s}}^{-1}$. 
These values of the bulk viscosity coefficient may be too large for what we expect today with realistic nuclear and particle physics models. However, they would still constitute the first upper limits on viscosity coefficients at these temperatures and densities obtained from astrophysical observations.}}

\section{Bulk viscosity in the post-merger system}
\label{sec:post}


{Having discussed that bulk 
viscosity might be challenging to extract from the inspiral,} we now turn to the post-merger evolution, where bulk viscous effects can be strongly enhanced.
{{In contrast, the estimated shear viscosity in the neutrino free streaming regime is much smaller \citep{Alford:2017rxf} so} we do not consider effects of microphysical shear viscosity.}

To obtain some indication of the likely importance of bulk viscous effects, we calculate some rough diagnostic indicators of the relevance of bulk viscosity in the background of
a representative simulation of an equal-mass 
binary neutron star merger with total mass $2.8\,M_{\odot}$
\citep{Most:2018eaw}.
This is the first time such an estimate of the likely relevance of bulk viscosity has been computed, and to obtain it we make certain simplifications. Firstly, the simulation itself does not include any bulk viscous effects; the purpose is to find out whether bulk viscous effects are significant enough to motivate such a fully self-consistent computation. Secondly, the EoS used to calculate the bulk viscosity is not the same as the EoS \citep{Dexheimer:2008ax} used in the simulation of the fluid dynamics of the merger. This is because the EoSs for which bulk viscosity calculations are currently available and those that have been used for simulations are different{, see also Sec. \ref{sec:bulk}} . However, we estimate the bulk viscous effects using three different EoSs, giving us an indication of the uncertainty associated with this mismatch.
The simulation presented here is modeled after the one presented in \citet{Most:2018eaw} and \citet{Most:2019onn}. We refer the reader to these works for details of the simulation parameters and to \citet{Most:2019kfe} for details on the numerical method.

We begin by describing the thermodynamic conditions present during the merger.
On first impact, depending on the EoS, temperatures up to about $80\, \rm MeV$ can be reached in parts of the newly formed neutron star \citep{Kastaun:2016yaf,Perego:2019adq}.
Bulk viscosity is a resonant phenomenon peaking at temperatures $\simeq 3\, \rm MeV$ \citep{Alford:2019qtm}, so we expect that
its effects
will be strongest in regions with temperatures in the MeV range, and suppressed as $\zeta\propto T^{-2}$ in regions where $T\gg 5\,{\rm MeV}$, which is also the regime where neutrinos become trapped in the star \citep{Roberts:2016mwj,Alford:2019kdw}. 
The thermodynamic evolution of the simulation presented here is illustrated in Fig. \ref{fig:temp}, which shows the temperatures, $T$, and number densities $n$ probed during the merger when using the CMF EoS \citep{Dexheimer:2008ax}.
We find that right at the moment of the initial collision, temperatures up to $T\simeq 40\, \rm MeV$ are reached at the merger interface (middle panel, top row) with about $T\simeq 30\, \rm MeV$ present during post-merger oscillations (right panel, top row). However, large parts of the star at densities $n\lesssim 2 n_{\rm sat}$,
remain cooler with $T\,<\, 10\, \rm MeV$.
In these regions the bulk viscosity for kHz oscillations reaches its resonant maximum and
our calculations will investigate whether it becomes strong enough to  damp the density oscillations at merger.

\subsection{Magnitude of bulk viscosity}
The 1\,kHz density oscillations are caused by the two neutron star cores repeatedly bouncing off each other before they eventually coalesce \citep{Takami:2014tva}.
{The damping process of these oscillations can be understood in terms of an effective bulk viscosity $\bar{\zeta}$ intrinsic to the merger, associated with gravitational wave emission.}
We can gain some insight into this behavior by considering the radial 
{displacement} $\left<r\right>$ of the two merging cores {from their center of mass}. 
Because we consider an equal-mass merger, the {two displacements are the same.}

{In the inviscid simulation that we are considering,} these oscillations are dampened over a dynamical time scale $\tau_{\rm damp}$ {that we can associate with}  an effective bulk viscosity $\bar{\zeta}$ operating at a frequency $\omega$ \citep{CerdaDuran:2009eh},
\begin{equation}
    \tau_{\mathrm{damp}}^{-1} = \frac{\omega^2\bar{\zeta}}{2\bar{\rho}\bar{c_s}^2},
\label{eqn:zeta_damp}
\end{equation}
where we have introduced an average rest-mass density $\bar{\rho} = \left(1.5\,\pm\, 0.5\right) \rho_{\rm sat}$ and corresponding sound speed $\bar{c}_s^2 \simeq \left(0.1\, \pm 0.05\right) c^2$.
{Although this expression has been formally derived for radial perturbations of a spherical fluid body, we can use it as a first approximation to associate the observed damping time scale $\tau_{\rm damp}$ with an average bulk viscosity $\bar{\zeta}$.}
The dynamics of the post-merger oscillations is shown in Fig. \ref{fig:pos}. 
We can see that the merging cores bounce at a frequency $f=\omega / \left(2 \pi\right) \simeq 1\, \rm kHz$, with a damping time scale {(due to gravitational radiation)} of $\tau_{\rm damp} \simeq \left( 30.4\,\pm\, 0.3\right)\, \rm ms$, although the oscillations will stop earlier when the two former stellar cores have merged into a single star.
Using ~\eqref{eqn:zeta_damp}, we can associate this with an effective bulk viscosity of 
\begin{align}
    \bar{\zeta} \approx \left(6\, \pm 4\right) \  {\times 10^{28}} \, \rm g/\left(cm\, s\right).
\label{eqn:barzeta}
\end{align}

For any microphysical bulk viscosity to affect the dynamics of the star and, hence, the gravitational wave emission, it should 
{be comparable to} the effective bulk viscosity $\bar{\zeta}$ estimated above.
{With \eqref{eqn:barzeta} and \eqref{eqn:Re2} as a reference scales, } 
we now evaluate the bulk viscosity on the background of an ideal hydrodynamical merger simulation. The center row of Fig. \ref{fig:temp} shows the values of the bulk viscosity for the ${\rm NL}\rho$ model. 
To provide a more quantitative assessment, we analyse the distribution of fluid elements in terms of the bulk viscosities they probe (see bottom row of Fig. \ref{fig:temp}). 
We can see that during the collision large parts of the star {probe} bulk viscosities between $10^{28}$ and $10^{30}\,\rm g/\left(\rm cm \, s\right)$.  This tells us that for this particular merger scenario, using the NL$\rho$ EoS to compute bulk viscosity, the {weak-interaction-driven bulk viscosity easily reaches values where it could outweigh} other damping mechanisms.

 We may quantify this more explicitly by computing the effective inverse Reynolds number \eqref{eqn:Re} associated with the damping of gravitational waves described above in terms of $\bar\zeta$ and $\bar\rho$. This can be done by assuming that 
the local fluid oscillations with frequency $\omega$ propagate with the sound speed $\bar{v}\simeq \bar{c}_s$, which in turn fixes the length scale $L \simeq \bar{c}_s/\omega$. For the flows observed in the inviscid simulation ($\tau_{\text{damp}}=30\,\text{ms}$, $\omega=2\pi\times (1\,\text{kHz})$), we find 
\begin{align}
    {\rm Re}^{-1}_{\text{inviscid}} = \frac{2}{\tau_{\rm damp}\, \omega} \approx 0.01\,.
    \label{eqn:Re2}
\end{align}
{It is important to highlight how our estimate of the post-merger ${\rm Re}^{-1}$ differs from \eqref{eqn:Pirho} that quantifies the Reynolds number in the inspiral via \eqref{eqn:chi_ref}. Since the compression in the post-merger oscillation is driven by the collisions of the two stars, and not by orbital decay, the $\left(v/c\right)^8$ scaling from the inspiral \eqref{eqn:Pirho} is not applicable ($\left(v/c\right)^8\rightarrow 1$ in the post-merger). Hence, the bulk viscous ratio $\chi$ can be significantly enhanced in the post-merger. } 
\begin{figure*}
    \centering
    \includegraphics[width=\textwidth]{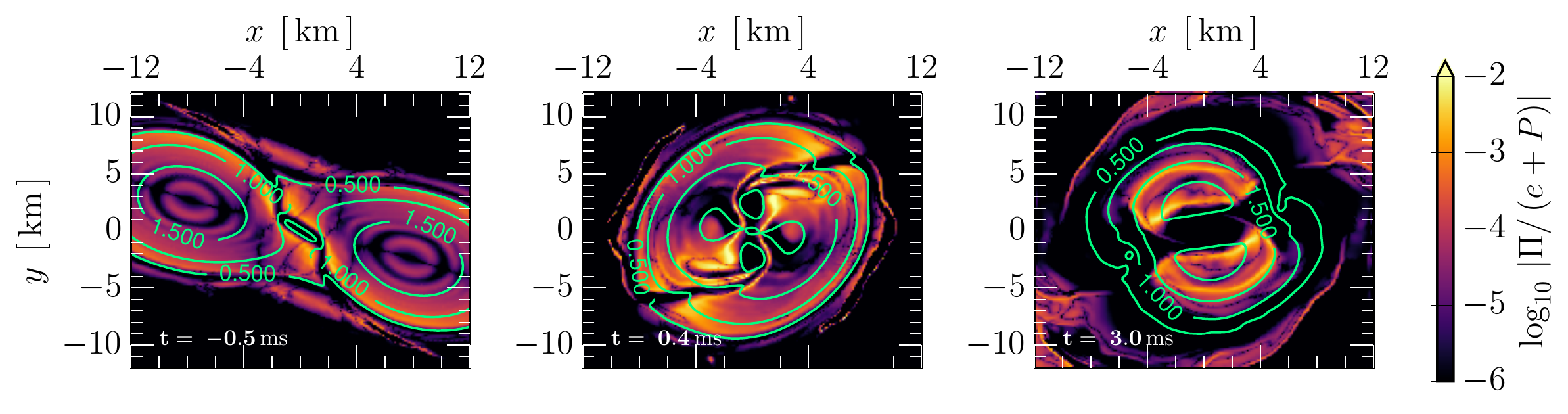}
    \caption{Relative importance of bulk viscosity in the late inspiral and early post-merger.
    {\it (Top)} Three representative times during the late inspiral and merger showing the relative fraction of the bulk scalar $\Pi$ to energy density $e$ and pressure $P$. The green lines are contours of baryon number density $n$ in units of nuclear saturation $n_{\rm sat}$. The bulk viscosity is computed using the $\rm NL\rho$ model.
    }
    \label{fig:chi}
\end{figure*}

\subsection{Bulk viscous ratio}
As a further test of the relevance of bulk viscosity, following our analysis of the inspiral \eqref{eqn:Pirho}, we now compute the bulk viscous ratio \eqref{eqn:bulk-viscous-ratio} coming from microphysical processes.

The simulation neglects bulk-viscous back reaction, so we estimate the bulk viscous pressure contribution by taking the Navier-Stokes limit \eqref{eqn:bulk_NS}.  If the bulk viscous ratio {(resulting from microphysical viscosity)} {approaches the magnitude} of the intrinsic inverse Reynolds number {associated with gravitational wave damping} \eqref{eqn:Re2}, then direct bulk viscous effects are considerable, which would invalidate our neglect of back-reaction. If the ratio is much less than \eqref{eqn:Re2} then the direct impact of bulk viscosity is small.
{During the post-merger oscillations motion in the compressional regime will reach $\bar{v}^2\simeq \bar{c}_s^2\simeq \left(0.1\,\pm\, 0.05\right)\, c^2$,
where the uncertainty range is taken to be  representative for the velocities attained at this stage.} 
Using \eqref{eqn:Re2} in \eqref{eqn:chi_ref} and the above velocity estimate, we can define a reference scale for the bulk viscous ratio as
\begin{align}
\chi_{\rm inviscid} \simeq \left(3.0\,\pm\, 1.5\right)\times 10^{-3}.
\label{eqn:chi_ref_inv}
\end{align}
{The estimate \eqref{eqn:chi_ref_inv} of the intrinsic bulk viscous ratio $\chi_{\rm inviscid}$ represents a natural reference scale for global compressional motion, as it is derived from the global damping time of post-merger oscillations. Although we find it useful to compare it to local compressional motion inside the stars, we caution that local oscillations at frequencies very different from $\omega$ might have other reference scales. A full determination of the relevant bulk viscous ratio necessary to affect the dynamics of the merger will ultimately require self-consistent viscous neutron star merger simulations. }

In Fig. \ref{fig:chi}, we show snapshots of the bulk viscous ratio $\chi$ at three times during the first few milliseconds of the merger, which is when the compression $-\nabla_\mu u^\mu$ is largest. In each snapshot we evaluate $\chi$ on the ideal fluid background of the simulation, using the $NL\rho$ equation of state.
Note that the EoS that we use to compute the bulk viscosity is not the same as the one used in the simulations. 
Our procedure is therefore not fully self-consistent, but it can give us an indication of the likely {magnitude of direct bulk viscous effects} 

We can see that right before the merger large parts of the star reach {$\chi \simeq 10^{-3}$}. This is perfectly consistent with the upper bound of the estimate performed in ~\eqref{eqn:Pirho}. 
During merger (middle and right panels of Fig. \ref{fig:chi}), large parts of the star are compressed and the relative bulk viscous pressure contribution locally reaches its maximum. At this time, a large fraction of the star has values of $\chi$ between $0.01$ and $0.1$, locally exceeding the bound for the intrinsic inverse Reynolds number \eqref{eqn:chi_ref_inv}, {indicating that microscopic bulk viscosity may play a significant role.}

\subsubsection{Density-weighted bulk-viscous ratio}
A rough way to characterize the direct effect of bulk viscosity on the entire merger system is via a density weighted average
\begin{align}
    \left<\chi\right> = \left< \frac{\Pi}{ \left(e+P\right)}\right> :=  \frac{ \int {\rm d}V\,\sqrt{\gamma}e \Pi/ \left(e+P\right)}
    {\int {\rm d}V\, \sqrt{\gamma}e }.
\end{align}
Here $\sqrt{\gamma}$ is the three-dimensional spatial volume element. Since high density regions affect the gravitational wave emission more strongly, $\left<\chi\right>$ provides an indication of the direct impact of bulk viscosity on gravitational wave emission at each instant during the merger.

We show the evolution of $\left<\chi \right>$ in Fig. \ref{fig:chi_max} for the three different models for weak interaction-driven bulk viscosity discussed in Sec.~\ref{sec:bulk}. 
The overall scale of $\langle\chi\rangle$ is {
around $(0.3 \text{-} 3)\times 10^{-4}$, not much smaller than the intrinsic inviscid value \eqref{eqn:chi_ref}, indicating that the direct
bulk viscous effect on gravitational wave emission may be noticeable. Moreover,} there are various
non-linear amplification mechanisms that could make bulk viscous effects even more important. 
For example, bulk viscous heating could bring cooler regions closer to the resonant maximum of bulk viscosity at $T\sim 4\,\text{MeV}$.
Nonlinear fluid mechanical effects could lead to effects on the amount of disk mass formation, dynamical mass ejection during the collision, as well as as the temperature distribution inside the remnant.
We note that bulk viscosity is also effective in shocks propagating from the merger remnant (right panel of Fig. \ref{fig:chi}). This opens up the tantalizing possibility of bulk viscosity to also affect dynamical mass ejection (see e.g. \citet{Abbott:2017wuw}).
While likely affecting only a small part of the material that will eventually become unbound and partake in the r-process nucleosynthesis  that gives rise to an electromagnetic afterglow (see e.g. \citet{Metzger:2019zeh} for a review), we cannot rule out the possibility of bulk viscous imprints on electromagnetic afterglows.

We also note that the variability across the different models shows how uncertainties in the nuclear physics can translate in to large differences in impact on the merger.
Focusing on the $\rm NL\rho$ model (red solid line), we can see that $\left<\chi\right>$ attains values of {$5\times 10^{-4}$} at merger and remains roughly constant on over a time scale $< 10\, \rm ms$ after merger.
In contrast, model $\rm BSR12$ (solid green line) reaches those maximum values in the inspiral but continuously declines in the post-merger.
These dramatic differences are related to the EoS-dependence of some of the nonlinearities discussed in the previous subsection: the bulk viscosity has a non-monotonic resonant dependence on temperature, with the resonant maximum depending on density and the EoS, as we saw in Fig.~\ref{fig:zeta}.

\subsubsection{Maximum bulk-viscous ratio}
\begin{figure}
    \centering
    \includegraphics[width=0.45\textwidth]{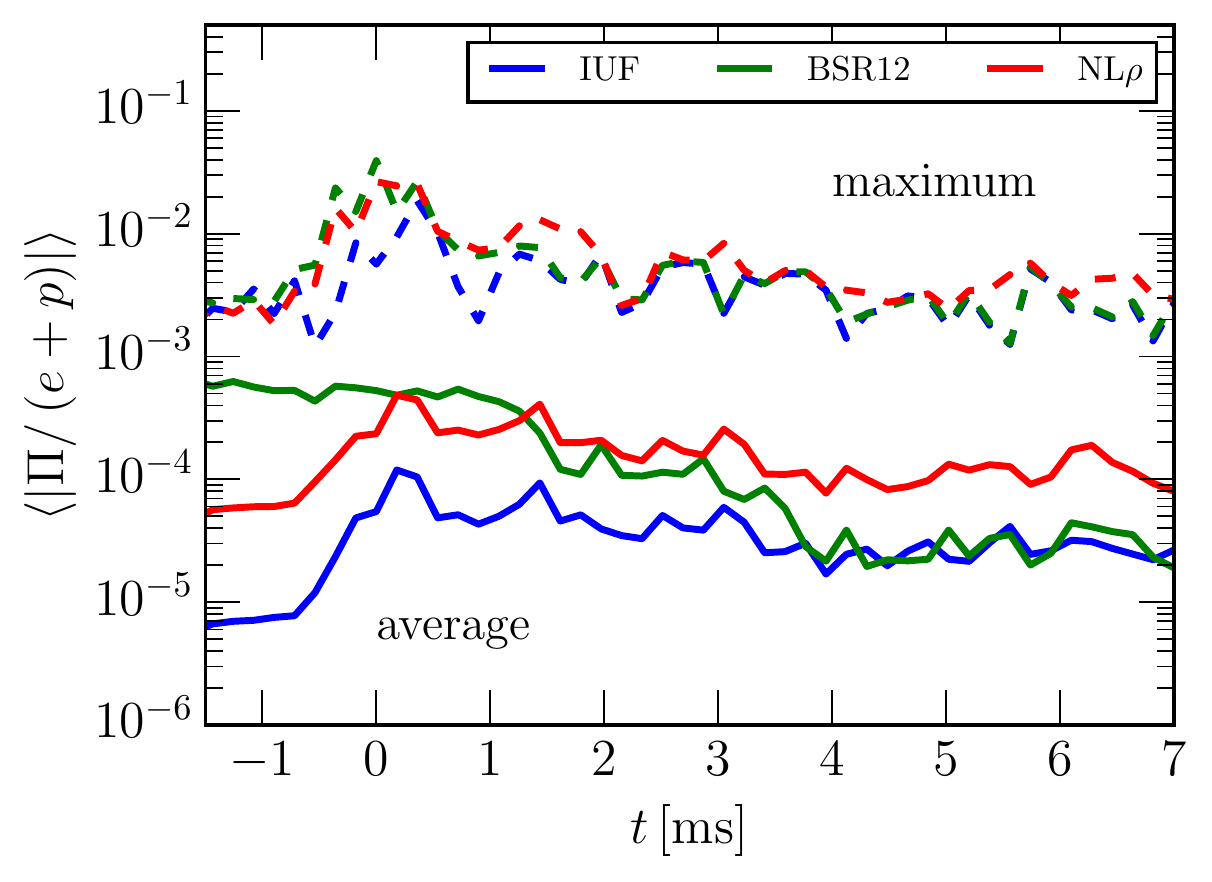}
    \caption{Bulk viscous ratio
    evaluated using three different nuclear matter models to compute the bulk viscosity. Shown are density weighted averages (solid lines) and maximum values (dashed lines). The time $t$ is defined relative to the time of merger.
    }
    \label{fig:chi_max}
\end{figure}

The maximum value of $\chi$ is of interest because it can be compared with other relativistic systems, see Sec.~\ref{sec:hic}. Its evolution is shown by the dashed lines in Fig. \ref{fig:chi_max}. Starting out at $10^{-3}$ in the inspiral, we can see that the maximum value of the bulk viscous ratio $\chi$ peaks around $5\%$ at the initial collision, and then drops to around $1\%$.
This behavior is independent of the EoS used to compute the bulk viscosity, with all of them leading to similar evolutions.
A comparison with heavy-ion collisions in Sec. \ref{sec:hic} suggests that such bulk viscous ratios are sufficient to affect dynamical evolution of a neutron star merger. 

\subsubsection{Bulk-viscous frequency shift}
It is interesting to note that one can estimate a global frequency associated with the appearance of bulk viscosity.
While previous analyses {(e.g. \citet{Alford:2017rxf,Alford:2019qtm}) }
have studied local oscillations of the fluid, we {can use our post-processing of a full merger simulation to investigate} the gravitational wave emission associated with the bulk component of the stress-energy tensor. More specifically, we use the quadrupole formula (see e.g. \citet{Baumgarte:2010ndz,Mueller:2012sv}) based on the energy component of the stress-energy tensor, i.e. $T^{00}_{\rm bulk} = \Pi \left(g^{00} + u^0 u^0\right)$, to estimate a gravitational wave strain $h_{\rm bulk}$ associated with the appearance of bulk viscosity.
Without proper backreaction, i.e. bulk viscous damping, onto the fluid flow such an estimate cannot be used to directly assess the impact of bulk viscosity on the post-merger spectrum of the gravitational wave signal \citep{Bauswein:2011tp,Bernuzzi:2012ci,Takami:2014zpa}. However, since the resonant behavior of the bulk viscosity, see Sec. \ref{sec:bulk}, might cause a shift between the frequency associated with compressional motion of the two stellar cores, it is interesting to look for deviations from the $f_2$ peak frequency \citep{Takami:2014zpa}. {We have calculated the gravitational wave emission and} we find that the frequency associated with the bulk viscous pressure contribution is shifted up by $\gtrsim 500\rm Hz$ compared to the $f_2$ frequency. We caution that since this result was obtained using only equatorial plane dynamics the error budget remains uncertain. Furthermore, fully back-reacted simulations using fully causal and strongly hyperbolic formulations of first-order dissipative effects \citep{Bemfica:2020zjp} will be needed to determine the actual impact of bulk viscosity on the post-merger evolution and gravitational wave emission.

\subsection{Comparison to heavy-ion collision dynamics}
\label{sec:hic}


To get a sense of the full impact of bulk viscosity in the evolution of mergers, including back-reaction and nonlinear effects, we can make a comparison with heavy-ion collisions where such calculations and measurements have already become the standard in the field for over a decade. We will argue that: (a)~the  values of the bulk viscous ratio $\chi$ that arise in heavy ion collisions are similar to those we estimated in mergers; (b)~bulk viscous effects in heavy-ion collisions are strong enough to observably affect the fluid-dynamical evolution.

Relativistic viscous hydrodynamic calculations have been employed in the field of heavy-ion collisions for well over a decade to describe the evolution of the quark-gluon plasma \citep{Romatschke:2017ejr}. The current state-of-the-art incorporates effects from shear and bulk viscosities in the  equations of motion following different prescriptions such as DNMR \citep{Denicol:2012cn}, BRSSS \cite{Baier:2007ix}, and anisotropic hydrodynamics \cite{Martinez:2010sc,Florkowski:2010cf,Bazow:2013ifa,Alqahtani:2017mhy,Alqahtani:2017jwl}. Also, the first numerical analyses of BDNK  \citep{Bemfica:2017wps,Kovtun:2019hdm,Bemfica:2019knx,Hoult:2020eho,Bemfica:2020zjp} have been performed in  \cite{Pandya:2021ief}. In fact, a significant effort is being expended in heavy-ion collisions towards understanding far-from-equilibrium hydrodynamics, as it pushes the very limits of causal fluid evolution \citep{Bemfica:2020xym,Plumberg:2021bme,Cheng:2021tnq} and creates the smallest droplet of fluid known to humanity \citep{Schenke:2021mxx}. 
\begin{figure}
    \centering
    \includegraphics[width=0.45\textwidth]{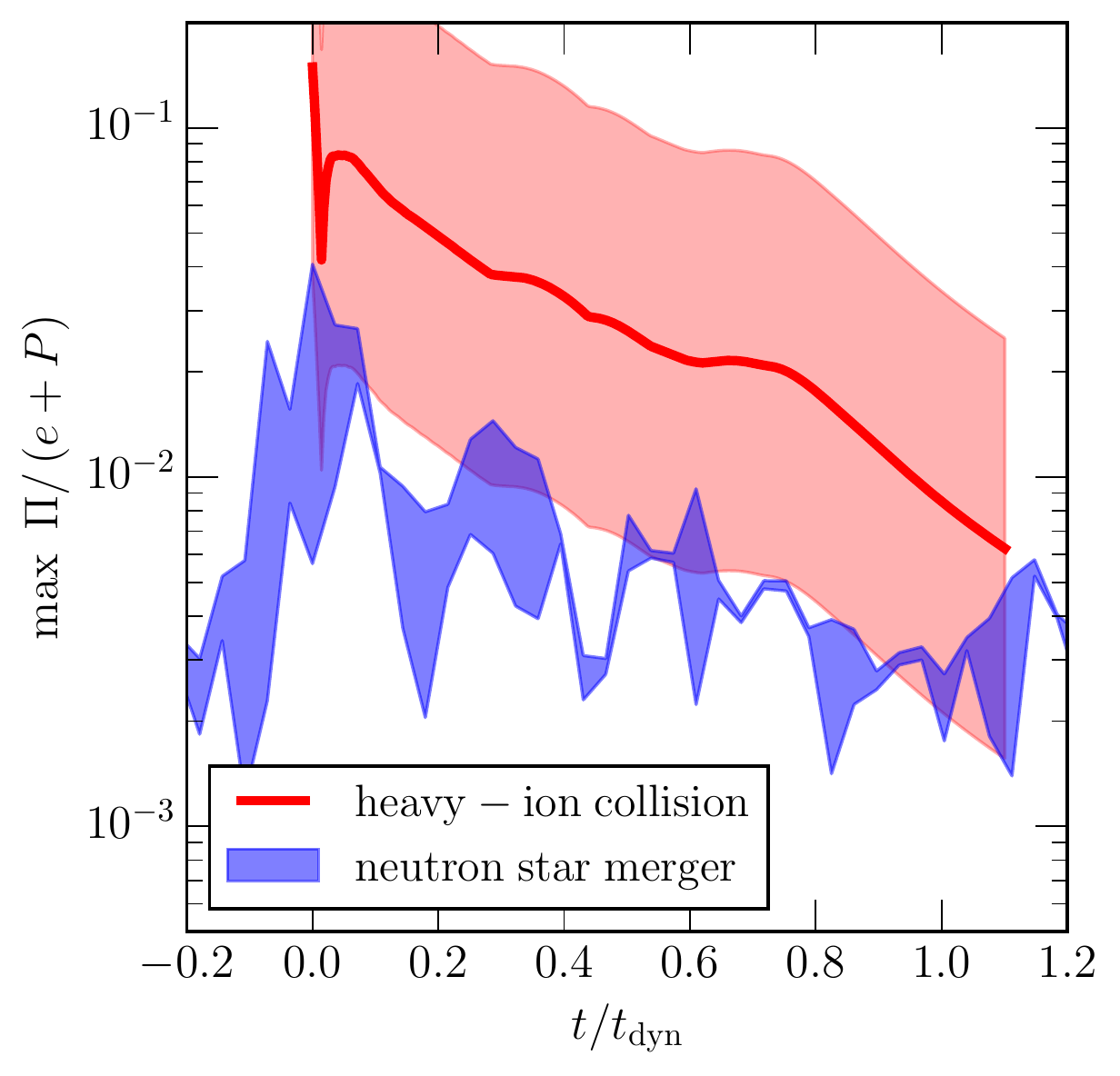}
    \caption{Comparison of the relative importance of bulk viscosity in heavy-ion collisions and neutron star mergers. Shown is the maximum value of the bulk viscous ratio $\chi = \Pi/\left(e+P\right)$. The evolution of the neutron star merger (blue band) includes the uncertainties of the bulk viscosity as shown in Fig.\ \ref{fig:chi}. The times have been normalized to a dynamical time scale $t_{\rm dyn}$ to ease a direct comparison.
    For the neutron star merger we choose $t_{\rm dyn} \simeq 5\,\rm ms$ corresponding to a characteristic damping timescale of early post-merger oscillations.
    In the case of the heavy-ion collision $t_{\rm dyn} \simeq 3 \times 10^{-20}\,\rm ms$ is the approximate lifetime of the system.
    }
    \label{fig:comparison}
\end{figure}

{There are obvious differences in  physical system size (heavy ion collisions have radii of $10^{-14}-10^{-15}$\,m) and in time scale: bulk viscosity in heavy ion collisions arises from thermal equilibration via strong interactions, operating on a timescale of $10^{-22}$ seconds}.
Despite these differences,
neutron star mergers and heavy-ion collisions should be connected as they rely on the same phase diagram of quantum chromodynamics \citep{Aryal:2020ocm,Dexheimer:2020zzs} and, at very low beam energies, there is overlap between the temperatures and densities achieved in heavy-ion collisions and the hot and dense matter formed in neutron star mergers \citep{Adamczewski-Musch:2019byl}. This motivates us to compare the order of magnitude of bulk viscous effects in both systems, since this has been studied already in great detail in the context of heavy-ion collisions \citep{Monnai:2009ad,Song:2009rh,Bozek:2009dw,Dusling:2011fd,Noronha-Hostler:2013gga, Ryu:2015vwa,Ryu:2017qzn}.

Heavy-ion collisions measured at RHIC and the LHC produce copious amounts of data leading to hundreds of experimental observables that can be used to constrain parameters in simulations.  This leads to tight constraints on transport coefficients (like bulk viscosity) and provides precise predictive power for new experiments \citep{Niemi:2015voa,Noronha-Hostler:2015dbi, Adam:2015ptt,Giacalone:2017dud,Shen:2020mgh,Acharya:2018ihu,Sirunyan:2019wqp,Aad:2019xmh}. More concretely, here we use the TRENTO+Free-stream+VISHNU+URQMD hydrodynamic setup \citep{Moreland:2014oya,Shen:2014vra,Petersen:2008dd}, which  has been used extensively in the field \citep{Bernhard:2019bmu,Shen:2020mgh,Bernhard:2019bmu}. The functional form of bulk viscosity in heavy-ion collisions employed here is extracted from a recent Bayesian analysis  \citep{Bernhard:2019bmu}  constrained by over 100 experimental observables -- for further details see \citet{Summerfield:2021oex}. 

Because the overall magnitude and temperature dependence of the bulk viscosity relevant to heavy-ion collisions is still not yet precisely known we have included an uncertainty band motivated by other model calculations such as \citep{Weller:2017tsr,Schenke:2020mbo} where the value of the bulk viscosity was not found using a Bayesian analysis. Additionally, we point out that in heavy-ion collisions there is also shear viscosity, which couples to bulk viscosity in the equations of motion \citep{Denicol:2012cn}. These coupling terms influence the order of magnitude of the values of $\Pi/(e+P)$, see \cite{Noronha-Hostler:2014dqa} for a detailed discussion in numerical simulations. In fact, some groups are even able to reproduce experimental data with no bulk viscosity and just shear viscosity \citep{Niemi:2015qia,Alba:2017hhe}. However, every Bayesian analysis performed in the field has shown a preference for a small bulk viscosity so we only consider simulations that implemented  finite values of $\zeta$ in the following.

In Fig.\ \ref{fig:comparison} the average contribution from the bulk viscosity, $\chi$, is shown over time in heavy-ion collisions compared to our order of magnitude estimates for neutron stars.  We find that the values are  comparable, albeit slightly smaller in the neutron star merger case. This indicates that bulk viscosity could play a role in the simulations of neutron star mergers as well. In fact, in heavy-ion collisions even small values of the bulk viscosity affect the evolution due to coupling terms to other transport coefficients \citep{Noronha-Hostler:2014dqa} and entropy production \citep{Dore:2020jye}. If the bulk viscosity in heavy-ion collisions is large than other observable consequences occur such as cavitation \citep{Torrieri:2008ip,Rajagopal:2009yw,Denicol:2015bpa,Byres:2019xld} may occur.
Similar effects may appear in neutron star mergers as well, although the effects of gravity may also lead to new and surprising consequences.
Fully general-relativistic viscous fluid dynamics simulations will be required to further investigate this point \citep{Shibata:2017xht}. 

\section{Conclusions}
In this paper we have assessed the possible impact of weak-interaction-driven bulk viscosity on neutron star mergers. 
Starting from {calculated} values of bulk viscosity from Urca processes, computed for a representative range of EoSs, we have estimated its impact on the inspiral gravitational wave signal and post-merger dynamics. 

\smallskip
\noindent \underline{Inspiral:} Initial calculations suggest that bulk viscosity {will not have a measureable impact on} gravitational waves emitted in the late stage inspiral of binary neutron stars.
We estimate that bulk viscosity enters at 4PN order. 
These viscous modifications will be enhanced by two powers of the compactness. However, since {from microphysical consideration} the coefficient of bulk viscosity $\zeta$ is expected to be {too small to be dynamically important}, and because of the PN suppression of these effects, their contribution to the gravitational wave signal is negligible unless the signal-to-noise ratios are absurdly large $( {\rm SNR} > 10^{5})$ 
or the microscopic estimate for $\zeta$ is too small by 3 orders of magnitude.

Since bulk viscosity {has a strong non-linear dependence on}  temperature, the impact on the gravitational wave signal will crucially depend on effects such as tidal heating and Urca processes in the inspiraling star. Given that these might critically affect the time before merger and hence the PN order at which they will become relevant, a more careful investigation will have to be performed to {{verify the qualitative inspiral conclusions discussed above.}}

\smallskip
\noindent \underline{Post merger:}
Based on an inviscid neutron star merger simulation, we estimated the direct impact of bulk viscosity on the merger dynamics in two ways, summarized below. 

First, we computed the bulk viscosity at each point in the merger region, and
found that the temperatures and densities probed during the merger are sufficient to produce bulk viscosities of $10^{28}$ to $10^{30}\, \rm g/\left( cm\ s\right)$ across large parts of the merger remnant. {This is comparable to the effective bulk viscosity estimated for inviscid damping mechanisms ($10^{28}$ to $10^{29}$ \,g/(cm\,s)) making bulk viscosity potentially comparable to those mechanisms.}

Next, we computed a figure of merit, the bulk viscous ratio \eqref{eqn:bulk-viscous-ratio}, {which is a proxy for the inverse Reynolds number of the flow}. It tells us that the bulk viscous contribution can be up to $5\%$ of the pressure shortly after the merger. Its density-weighted average across the merger region is  {around $(0.3 \,\text{-}\, 3)\times 10^{-4}$}, with significant variability depending on the assumed EoS. 
This is {not much smaller than the effective inverse Reynolds number of inviscid damping mechanisms, $\left(3.0\,\pm\, 1.5\right)\times 10^{-3}$}, indicating again that the {bulk viscosity could be a non-negligible contribution to the damping.}

These measurements of the direct impact of bulk viscosity should be understood as a conservative estimate, since there are additional factors that could lead to a greater impact. We only studied one particular merger scenario (equal mass merger with one EoS), and it is reasonable to expect that other configurations or EoSs will lead to different results, some of which will show greater impact.
The presence of sizeable bulk viscous ratios in outward-moving lower density regions during the merger indicates that early mass ejection might be affected by the presence of bulk viscosity as well.
{Additionally, there could be amplification of bulk viscous effects via nonlinear dynamics.
For example, bulk-viscous heating that would locally increase the temperature combined with the non-monotonic dependence of bulk viscosity on temperature, or nonlinear fluid dynamics.
To illustrate this possibility, we compared our results}
with the conditions found in relativistic heavy-ion collisions, {which have been studied in great detail, including back-reaction and non-linear effects.} {We find that the bulk-viscous ratios are not very different, giving a further indication that bulk viscosity could}  affect the flow structure and, hence, non-linearly affect the ejection of mass, the life-time of the hypermassive neutron star, and the post-merger gravitational wave signal.

Overall, some of these effects might be comparable to those resulting from the inclusion of shear viscosity. In fact, simulations modeling the effects of under-resolved magneto-turbulence in the merger have found \citep{Shibata:2017xht,Radice:2017zta} potentially rapid suppression of the gravitational wave emission.
We anticipate that bulk viscosity  could  affect the post-merger gravitational wave emission in a similar way, as the oscillation of the two stellar cores after merger could be rapidly damped, leading to a faster decay of the signal.

In performing these estimates, we have relied on several assumptions.
We {used an inviscid simulation, neglecting} the back reaction of bulk viscosity on the fluid flow.
This means that we are likely overestimating the amount of compression $-\nabla_\mu u^\mu>0$. At the same time, the lack of back reaction also neglects the change in temperature and flow structure due to bulk viscosity, although it is difficult to predict whether this would drive the fluid into or out of the resonant bulk viscous regime.
As briefly noted above, we have considered only a single neutron star merger simulation for a given EoS. It seems likely that some other systems with different EoS, total mass, or mass ratio will provide conditions where bulk viscosity has different
effects.

Our overall conclusion is that there is good reason to pursue more careful investigations of the impact of bulk viscosity, exploring a range of merger scenarios and using
simulations that, unlike the one used here, fully incorporate bulk viscosity, including its back-reaction on the fluid flow. In addition, it will be crucial to explore the combination of  finite-temperature effects in the EoS \citep{Raithel:2021hye} and temperature dependence of bulk viscosity, as the thermodynamic conditions are crucial to the understanding of bulk viscous effects.
Ultimately, only such self-consistent numerical relativity simulations of neutron star mergers with bulk viscosity using strongly hyperbolic, causal dissipative hydrodynamics \citep{Bemfica:2019cop,Bemfica:2020zjp} will be able to fully clarify the role of bulk viscosity in the post-merger evolution. 

Looking further ahead, one could
investigate possible signatures of exotic degrees of freedom such as hyperons \citep{Alford:2020pld}, quarks \citep{Alford:2006gy,Alford:2008pb,Bierkandt:2011zp,Harutyunyan:2017ieu}, nuclear pasta \cite{Yakovlev:2018jia,Lin:2020nxy}, and quark-hadron mixed phases \citep{Drago:2003wg} and to take into account neutrino trapping which is expected to become important once the temperatures in the merger remnant increase beyond a few $\rm MeV$. 
Finally, we have assumed the ``subthermal'' limit of linear response, but large amplitude oscillations may experience ``suprathermal'' bulk viscosity \citep{Madsen:1992sx,Reisenegger:1994be,Alford:2010gw}, the impact of which has not yet been estimated outside of the inspiral phase \citep{Arras:2018fxj}.

\section*{Acknowledgements}
ERM thanks Carolyn Raithel for insightful discussions on finite-temperature effects.
ERM gratefully acknowledges support from a joint fellowship
at the Princeton Center for Theoretical Science, the Princeton Gravity
Initiative and the Institute for Advanced Study.
SPH is supported by the U.~S. Department of Energy grant DE-FG02-00ER41132 as well as the National Science Foundation grant No.~PHY-1430152 (JINA Center for the Evolution of the Elements).  JN is partially supported by the U.S.~Department of Energy, Office of Science, Office for Nuclear Physics under Award No.~DE-SC0021301. JNH and CP acknowledge the support from the US-DOE Nuclear Science Grant No. DE-SC0020633. MGA is partially supported by the U.S.~Department of Energy, Office of Science, Office of Nuclear Physics under Award No.~\#DE-FG02-05ER41375. FP acknowledges support from NSF Grant No. PHY-1912171, the Simons Foundation, and the Canadian Institute For Advanced Research (CIFAR).
HW acknowledges financial support provided by the NSF Grant No. OAC-2004879.
HW thanks the Albert-Einstein Institute (AEI) Potsdam for kind hospitality while finishing this work.
The simulation presented in this article was performed on computational resources managed and supported by Princeton Research Computing, a consortium of groups including the Princeton Institute for Computational Science and Engineering (PICSciE) and the Office of Information Technology's High Performance Computing Center and Visualization Laboratory at Princeton University.
{Part of this work was performed at the Aspen Center for Physics, which is supported by National Science Foundation grant PHY-1607611. The participation of ERM at the Aspen Center for Physics was supported by the Simons Foundation.}

\section*{Data Availability}
Data is available upon reasonable request from the authors.
%

%
\bibliographystyle{mnras}
\bibliography{inspire,non_inspire} 

\begin{thebibliography}{}
\makeatletter
\relax
\def\mn@urlcharsother{\let\do\@makeother \do\$\do\&\do\#\do\^\do\_\do\%\do\~}
\def\mn@doi{\begingroup\mn@urlcharsother \@ifnextchar [ {\mn@doi@}
  {\mn@doi@[]}}
\def\mn@doi@[#1]#2{\def\@tempa{#1}\ifx\@tempa\@empty \href
  {http://dx.doi.org/#2} {doi:#2}\else \href {http://dx.doi.org/#2} {#1}\fi
  \endgroup}
\def\mn@eprint#1#2{\mn@eprint@#1:#2::\@nil}
\def\mn@eprint@arXiv#1{\href {http://arxiv.org/abs/#1} {{\tt arXiv:#1}}}
\def\mn@eprint@dblp#1{\href {http://dblp.uni-trier.de/rec/bibtex/#1.xml}
  {dblp:#1}}
\def\mn@eprint@#1:#2:#3:#4\@nil{\def\@tempa {#1}\def\@tempb {#2}\def\@tempc
  {#3}\ifx \@tempc \@empty \let \@tempc \@tempb \let \@tempb \@tempa \fi \ifx
  \@tempb \@empty \def\@tempb {arXiv}\fi \@ifundefined
  {mn@eprint@\@tempb}{\@tempb:\@tempc}{\expandafter \expandafter \csname
  mn@eprint@\@tempb\endcsname \expandafter{\@tempc}}}

\bibitem[\protect\citeauthoryear{Aad et~al.}{Aad et~al.}{2020}]{Aad:2019xmh}
Aad G.,  et~al., 2020, \mn@doi [Phys. Rev. C] {10.1103/PhysRevC.101.024906},
  101, 024906

\bibitem[\protect\citeauthoryear{Abbott et~al.}{Abbott
  et~al.}{2017a}]{TheLIGOScientific:2017qsa}
Abbott B.~P.,  et~al., 2017a, \mn@doi [Phys. Rev. Lett.]
  {10.1103/PhysRevLett.119.161101}, 119, 161101

\bibitem[\protect\citeauthoryear{Abbott et~al.}{Abbott
  et~al.}{2017b}]{GBM:2017lvd}
Abbott B.~P.,  et~al., 2017b, \mn@doi [Astrophys. J. Lett.]
  {10.3847/2041-8213/aa91c9}, 848, L12

\bibitem[\protect\citeauthoryear{Abbott et~al.}{Abbott
  et~al.}{2017c}]{Abbott:2017wuw}
Abbott B.~P.,  et~al., 2017c, \mn@doi [Astrophys. J. Lett.]
  {10.3847/2041-8213/aa9478}, 850, L39

\bibitem[\protect\citeauthoryear{Abbott et~al.}{Abbott
  et~al.}{2018}]{Abbott:2018exr}
Abbott B.~P.,  et~al., 2018, \mn@doi [Phys. Rev. Lett.]
  {10.1103/PhysRevLett.121.161101}, 121, 161101

\bibitem[\protect\citeauthoryear{Abbott et~al.}{Abbott
  et~al.}{2020}]{Abbott:2020uma}
Abbott B.~P.,  et~al., 2020, \mn@doi [Astrophys. J. Lett.]
  {10.3847/2041-8213/ab75f5}, 892, L3

\bibitem[\protect\citeauthoryear{Acharya et~al.}{Acharya
  et~al.}{2018}]{Acharya:2018ihu}
Acharya S.,  et~al., 2018, \mn@doi [Phys. Lett. B]
  {10.1016/j.physletb.2018.06.059}, 784, 82

\bibitem[\protect\citeauthoryear{Adam et~al.}{Adam et~al.}{2016}]{Adam:2015ptt}
Adam J.,  et~al., 2016, \mn@doi [Phys. Rev. Lett.]
  {10.1103/PhysRevLett.116.222302}, 116, 222302

\bibitem[\protect\citeauthoryear{Adamczewski-Musch et~al.}{Adamczewski-Musch
  et~al.}{2019}]{Adamczewski-Musch:2019byl}
Adamczewski-Musch J.,  et~al., 2019, \mn@doi [Nature Phys.]
  {10.1038/s41567-019-0583-8}, 15, 1040

\bibitem[\protect\citeauthoryear{Alba, Mantovani~Sarti, Noronha,
  Noronha-Hostler, Parotto, Portillo~Vazquez  \& Ratti}{Alba
  et~al.}{2018}]{Alba:2017hhe}
Alba P.,  Mantovani~Sarti V.,  Noronha J.,  Noronha-Hostler J.,  Parotto P.,
  Portillo~Vazquez I.,   Ratti C.,  2018, \mn@doi [Phys. Rev. C]
  {10.1103/PhysRevC.98.034909}, 98, 034909

\bibitem[\protect\citeauthoryear{Alford \& Haber}{Alford \&
  Haber}{2021}]{Alford:2020pld}
Alford M.~G.,  Haber A.,  2021, \mn@doi [Phys. Rev. C]
  {10.1103/PhysRevC.103.045810}, 103, 045810

\bibitem[\protect\citeauthoryear{Alford \& Harris}{Alford \&
  Harris}{2018}]{Alford:2018lhf}
Alford M.~G.,  Harris S.~P.,  2018, \mn@doi [Phys. Rev. C]
  {10.1103/PhysRevC.98.065806}, 98, 065806

\bibitem[\protect\citeauthoryear{Alford \& Harris}{Alford \&
  Harris}{2019}]{Alford:2019qtm}
Alford M.~G.,  Harris S.~P.,  2019, \mn@doi [Phys. Rev. C]
  {10.1103/PhysRevC.100.035803}, 100, 035803

\bibitem[\protect\citeauthoryear{Alford \& Schmitt}{Alford \&
  Schmitt}{2007}]{Alford:2006gy}
Alford M.~G.,  Schmitt A.,  2007, \mn@doi [J. Phys. G]
  {10.1088/0954-3899/34/1/005}, 34, 67

\bibitem[\protect\citeauthoryear{Alford, Braby  \& Schmitt}{Alford
  et~al.}{2008}]{Alford:2008pb}
Alford M.~G.,  Braby M.,   Schmitt A.,  2008, \mn@doi [J. Phys. G]
  {10.1088/0954-3899/35/11/115007}, 35, 115007

\bibitem[\protect\citeauthoryear{Alford, Mahmoodifar  \& Schwenzer}{Alford
  et~al.}{2010}]{Alford:2010gw}
Alford M.~G.,  Mahmoodifar S.,   Schwenzer K.,  2010, \mn@doi [J. Phys. G]
  {10.1088/0954-3899/37/12/125202}, 37, 125202

\bibitem[\protect\citeauthoryear{Alford, Bovard, Hanauske, Rezzolla  \&
  Schwenzer}{Alford et~al.}{2018}]{Alford:2017rxf}
Alford M.~G.,  Bovard L.,  Hanauske M.,  Rezzolla L.,   Schwenzer K.,  2018,
  \mn@doi [Phys. Rev. Lett.] {10.1103/PhysRevLett.120.041101}, 120, 041101

\bibitem[\protect\citeauthoryear{Alford, Harutyunyan  \& Sedrakian}{Alford
  et~al.}{2019}]{Alford:2019kdw}
Alford M.,  Harutyunyan A.,   Sedrakian A.,  2019, \mn@doi [Phys. Rev. D]
  {10.1103/PhysRevD.100.103021}, 100, 103021

\bibitem[\protect\citeauthoryear{Alford, Harutyunyan  \& Sedrakian}{Alford
  et~al.}{2020}]{Alford:2020lla}
Alford M.,  Harutyunyan A.,   Sedrakian A.,  2020, \mn@doi [Particles]
  {10.3390/particles3020034}, 3, 500

\bibitem[\protect\citeauthoryear{Alqahtani, Nopoush, Ryblewski  \&
  Strickland}{Alqahtani et~al.}{2017}]{Alqahtani:2017jwl}
Alqahtani M.,  Nopoush M.,  Ryblewski R.,   Strickland M.,  2017, \mn@doi
  [Phys. Rev. Lett.] {10.1103/PhysRevLett.119.042301}, 119, 042301

\bibitem[\protect\citeauthoryear{Alqahtani, Nopoush  \& Strickland}{Alqahtani
  et~al.}{2018}]{Alqahtani:2017mhy}
Alqahtani M.,  Nopoush M.,   Strickland M.,  2018, \mn@doi [Prog. Part. Nucl.
  Phys.] {10.1016/j.ppnp.2018.05.004}, 101, 204

\bibitem[\protect\citeauthoryear{Annala, Gorda, Kurkela  \& Vuorinen}{Annala
  et~al.}{2018}]{Annala:2017llu}
Annala E.,  Gorda T.,  Kurkela A.,   Vuorinen A.,  2018, \mn@doi [Phys. Rev.
  Lett.] {10.1103/PhysRevLett.120.172703}, 120, 172703

\bibitem[\protect\citeauthoryear{Arras \& Weinberg}{Arras \&
  Weinberg}{2019}]{Arras:2018fxj}
Arras P.,  Weinberg N.~N.,  2019, \mn@doi [Mon. Not. Roy. Astron. Soc.]
  {10.1093/mnras/stz880}, 486, 1424

\bibitem[\protect\citeauthoryear{Aryal, Constantinou, Farias  \&
  Dexheimer}{Aryal et~al.}{2020}]{Aryal:2020ocm}
Aryal K.,  Constantinou C.,  Farias R. L.~S.,   Dexheimer V.,  2020, \mn@doi
  [Phys. Rev. D] {10.1103/PhysRevD.102.076016}, 102, 076016

\bibitem[\protect\citeauthoryear{Baier, Romatschke, Son, Starinets  \&
  Stephanov}{Baier et~al.}{2008}]{Baier:2007ix}
Baier R.,  Romatschke P.,  Son D.~T.,  Starinets A.~O.,   Stephanov M.~A.,
  2008, \mn@doi [JHEP] {10.1088/1126-6708/2008/04/100}, 04, 100

\bibitem[\protect\citeauthoryear{Baiotti}{Baiotti}{2019}]{Baiotti:2019sew}
Baiotti L.,  2019, \mn@doi [Prog. Part. Nucl. Phys.]
  {10.1016/j.ppnp.2019.103714}, 109, 103714

\bibitem[\protect\citeauthoryear{Baumgarte \& Shapiro}{Baumgarte \&
  Shapiro}{2010}]{Baumgarte:2010ndz}
Baumgarte T.~W.,  Shapiro S.~L.,  2010, {Numerical Relativity: Solving
  Einstein's Equations on the Computer}.
Cambridge University Press, \mn@doi{10.1017/CBO9781139193344}

\bibitem[\protect\citeauthoryear{Bauswein \& Janka}{Bauswein \&
  Janka}{2012}]{Bauswein:2011tp}
Bauswein A.,  Janka H.~T.,  2012, \mn@doi [Phys. Rev. Lett.]
  {10.1103/PhysRevLett.108.011101}, 108, 011101

\bibitem[\protect\citeauthoryear{Bauswein, Just, Janka  \&
  Stergioulas}{Bauswein et~al.}{2017}]{Bauswein:2017vtn}
Bauswein A.,  Just O.,  Janka H.-T.,   Stergioulas N.,  2017, \mn@doi
  [Astrophys. J. Lett.] {10.3847/2041-8213/aa9994}, 850, L34

\bibitem[\protect\citeauthoryear{Bauswein, Bastian, Blaschke, Chatziioannou,
  Clark, Fischer  \& Oertel}{Bauswein et~al.}{2019}]{Bauswein:2018bma}
Bauswein A.,  Bastian N.-U.~F.,  Blaschke D.~B.,  Chatziioannou K.,  Clark
  J.~A.,  Fischer T.,   Oertel M.,  2019, \mn@doi [Phys. Rev. Lett.]
  {10.1103/PhysRevLett.122.061102}, 122, 061102

\bibitem[\protect\citeauthoryear{Bazow, Heinz  \& Strickland}{Bazow
  et~al.}{2014}]{Bazow:2013ifa}
Bazow D.,  Heinz U.~W.,   Strickland M.,  2014, \mn@doi [Phys. Rev. C]
  {10.1103/PhysRevC.90.054910}, 90, 054910

\bibitem[\protect\citeauthoryear{Beloin, Han, Steiner  \& Odbadrakh}{Beloin
  et~al.}{2019}]{Beloin:2018fyp}
Beloin S.,  Han S.,  Steiner A.~W.,   Odbadrakh K.,  2019, \mn@doi [Phys. Rev.
  C] {10.1103/PhysRevC.100.055801}, 100, 055801

\bibitem[\protect\citeauthoryear{Bemfica, Disconzi  \& Noronha}{Bemfica
  et~al.}{2018}]{Bemfica:2017wps}
Bemfica F.~S.,  Disconzi M.~M.,   Noronha J.,  2018, \mn@doi [Phys. Rev. D]
  {10.1103/PhysRevD.98.104064}, 98, 104064

\bibitem[\protect\citeauthoryear{Bemfica, Disconzi  \& Noronha}{Bemfica
  et~al.}{2019a}]{Bemfica:2019knx}
Bemfica F.~S.,  Disconzi M.~M.,   Noronha J.,  2019a, \mn@doi [Phys. Rev. D]
  {10.1103/PhysRevD.100.104020}, 100, 104020

\bibitem[\protect\citeauthoryear{Bemfica, Disconzi  \& Noronha}{Bemfica
  et~al.}{2019b}]{Bemfica:2019cop}
Bemfica F.~S.,  Disconzi M.~M.,   Noronha J.,  2019b, \mn@doi [Phys. Rev.
  Lett.] {10.1103/PhysRevLett.122.221602}, 122, 221602

\bibitem[\protect\citeauthoryear{Bemfica, Disconzi  \& Noronha}{Bemfica
  et~al.}{2020a}]{Bemfica:2020zjp}
Bemfica F.~S.,  Disconzi M.~M.,   Noronha J.,  2020a

\bibitem[\protect\citeauthoryear{Bemfica, Disconzi, Hoang, Noronha  \&
  Radosz}{Bemfica et~al.}{2020b}]{Bemfica:2020xym}
Bemfica F.~S.,  Disconzi M.~M.,  Hoang V.,  Noronha J.,   Radosz M.,  2020b

\bibitem[\protect\citeauthoryear{Bernhard, Moreland  \& Bass}{Bernhard
  et~al.}{2019}]{Bernhard:2019bmu}
Bernhard J.~E.,  Moreland J.~S.,   Bass S.~A.,  2019, \mn@doi [Nature Phys.]
  {10.1038/s41567-019-0611-8}, 15, 1113

\bibitem[\protect\citeauthoryear{Bernuzzi, Nagar, Thierfelder  \&
  Brugmann}{Bernuzzi et~al.}{2012}]{Bernuzzi:2012ci}
Bernuzzi S.,  Nagar A.,  Thierfelder M.,   Brugmann B.,  2012, \mn@doi [Phys.
  Rev. D] {10.1103/PhysRevD.86.044030}, 86, 044030

\bibitem[\protect\citeauthoryear{Beznogov \& Yakovlev}{Beznogov \&
  Yakovlev}{2015}]{Beznogov:2015ewa}
Beznogov M.~V.,  Yakovlev D.~G.,  2015, \mn@doi [Mon. Not. Roy. Astron. Soc.]
  {10.1093/mnras/stv1293}, 452, 540

\bibitem[\protect\citeauthoryear{Bierkandt \& Manuel}{Bierkandt \&
  Manuel}{2011}]{Bierkandt:2011zp}
Bierkandt R.,  Manuel C.,  2011, \mn@doi [Phys. Rev. D]
  {10.1103/PhysRevD.84.023004}, 84, 023004

\bibitem[\protect\citeauthoryear{Blacker, Bastian, Bauswein, Blaschke, Fischer,
  Oertel, Soultanis  \& Typel}{Blacker et~al.}{2020}]{Blacker:2020nlq}
Blacker S.,  Bastian N.-U.~F.,  Bauswein A.,  Blaschke D.~B.,  Fischer T.,
  Oertel M.,  Soultanis T.,   Typel S.,  2020, \mn@doi [Phys. Rev. D]
  {10.1103/PhysRevD.102.123023}, 102, 123023

\bibitem[\protect\citeauthoryear{Blanchet}{Blanchet}{2014}]{Blanchet:2013haa}
Blanchet L.,  2014, \mn@doi [Living Rev. Rel.] {10.12942/lrr-2014-2}, 17, 2

\bibitem[\protect\citeauthoryear{Bose, Chakravarti, Rezzolla, Sathyaprakash  \&
  Takami}{Bose et~al.}{2018}]{Bose:2017jvk}
Bose S.,  Chakravarti K.,  Rezzolla L.,  Sathyaprakash B.~S.,   Takami K.,
  2018, \mn@doi [Phys. Rev. Lett.] {10.1103/PhysRevLett.120.031102}, 120,
  031102

\bibitem[\protect\citeauthoryear{Bozek}{Bozek}{2010}]{Bozek:2009dw}
Bozek P.,  2010, \mn@doi [Phys. Rev. C] {10.1103/PhysRevC.81.034909}, 81,
  034909

\bibitem[\protect\citeauthoryear{Brown, Cumming, Fattoyev, Horowitz, Page  \&
  Reddy}{Brown et~al.}{2018}]{Brown:2017gxd}
Brown E.~F.,  Cumming A.,  Fattoyev F.~J.,  Horowitz C.~J.,  Page D.,   Reddy
  S.,  2018, \mn@doi [Phys. Rev. Lett.] {10.1103/PhysRevLett.120.182701}, 120,
  182701

\bibitem[\protect\citeauthoryear{Byres, Lim, McGinn, Ouellette  \& Nagle}{Byres
  et~al.}{2020}]{Byres:2019xld}
Byres M.,  Lim S.~H.,  McGinn C.,  Ouellette J.,   Nagle J.~L.,  2020, \mn@doi
  [Phys. Rev. C] {10.1103/PhysRevC.101.044902}, 101, 044902

\bibitem[\protect\citeauthoryear{Cerda-Duran}{Cerda-Duran}{2010}]{CerdaDuran:2009eh}
Cerda-Duran P.,  2010, \mn@doi [Class. Quant. Grav.]
  {10.1088/0264-9381/27/20/205012}, 27, 205012

\bibitem[\protect\citeauthoryear{Chabanov, Rezzolla  \& Rischke}{Chabanov
  et~al.}{2021}]{Chabanov:2021dee}
Chabanov M.,  Rezzolla L.,   Rischke D.~H.,  2021, ] {10.1093/mnras/stab1384}

\bibitem[\protect\citeauthoryear{Cheng \& Shen}{Cheng \&
  Shen}{2021}]{Cheng:2021tnq}
Cheng C.,  Shen C.,  2021

\bibitem[\protect\citeauthoryear{{Cutler}, {Lindblom}  \& {Splinter}}{{Cutler}
  et~al.}{1990}]{1990ApJ...363..603C}
{Cutler} C.,  {Lindblom} L.,   {Splinter} R.~J.,  1990, \mn@doi [\apj]
  {10.1086/169370}, \href
  {https://ui.adsabs.harvard.edu/abs/1990ApJ...363..603C} {363, 603}

\bibitem[\protect\citeauthoryear{Denicol \& Noronha}{Denicol \&
  Noronha}{2020}]{Denicol:2019lio}
Denicol G.~S.,  Noronha J.,  2020, \mn@doi [Phys. Rev. Lett.]
  {10.1103/PhysRevLett.124.152301}, 124, 152301

\bibitem[\protect\citeauthoryear{Denicol, Niemi, Molnar  \& Rischke}{Denicol
  et~al.}{2012}]{Denicol:2012cn}
Denicol G.~S.,  Niemi H.,  Molnar E.,   Rischke D.~H.,  2012, \mn@doi [Phys.
  Rev. D] {10.1103/PhysRevD.85.114047}, 85, 114047

\bibitem[\protect\citeauthoryear{Denicol, Gale  \& Jeon}{Denicol
  et~al.}{2015}]{Denicol:2015bpa}
Denicol G.~S.,  Gale C.,   Jeon S.,  2015, \mn@doi [PoS] {10.22323/1.217.0033},
  CPOD2014, 033

\bibitem[\protect\citeauthoryear{Dexheimer \& Schramm}{Dexheimer \&
  Schramm}{2008}]{Dexheimer:2008ax}
Dexheimer V.,  Schramm S.,  2008, \mn@doi [Astrophys. J.] {10.1086/589735},
  683, 943

\bibitem[\protect\citeauthoryear{Dexheimer, Noronha, Noronha-Hostler, Ratti  \&
  Yunes}{Dexheimer et~al.}{2020}]{Dexheimer:2020zzs}
Dexheimer V.,  Noronha J.,  Noronha-Hostler J.,  Ratti C.,   Yunes N.,  2020

\bibitem[\protect\citeauthoryear{Dhiman, Kumar  \& Agrawal}{Dhiman
  et~al.}{2007}]{Dhiman:2007ck}
Dhiman S.~K.,  Kumar R.,   Agrawal B.~K.,  2007, \mn@doi [Phys. Rev. C]
  {10.1103/PhysRevC.76.045801}, 76, 045801

\bibitem[\protect\citeauthoryear{Dore, Noronha-Hostler  \& McLaughlin}{Dore
  et~al.}{2020}]{Dore:2020jye}
Dore T.,  Noronha-Hostler J.,   McLaughlin E.,  2020, \mn@doi [Phys. Rev. D]
  {10.1103/PhysRevD.102.074017}, 102, 074017

\bibitem[\protect\citeauthoryear{Drago, Lavagno  \& Pagliara}{Drago
  et~al.}{2005}]{Drago:2003wg}
Drago A.,  Lavagno A.,   Pagliara G.,  2005, \mn@doi [Phys. Rev. D]
  {10.1103/PhysRevD.71.103004}, 71, 103004

\bibitem[\protect\citeauthoryear{Duez et~al.,}{Duez
  et~al.}{2020}]{Duez:2020lgq}
Duez M.~D.,  et~al., 2020, \mn@doi [Phys. Rev. D]
  {10.1103/PhysRevD.102.104050}, 102, 104050

\bibitem[\protect\citeauthoryear{Dusling \& Sch\"afer}{Dusling \&
  Sch\"afer}{2012}]{Dusling:2011fd}
Dusling K.,  Sch\"afer T.,  2012, \mn@doi [Phys. Rev. C]
  {10.1103/PhysRevC.85.044909}, 85, 044909

\bibitem[\protect\citeauthoryear{Dutra et~al.,}{Dutra
  et~al.}{2014}]{Dutra:2014qga}
Dutra M.,  et~al., 2014, \mn@doi [Phys. Rev. C] {10.1103/PhysRevC.90.055203},
  90, 055203

\bibitem[\protect\citeauthoryear{Endrizzi et~al.,}{Endrizzi
  et~al.}{2020}]{Endrizzi:2019trv}
Endrizzi A.,  et~al., 2020, \mn@doi [Eur. Phys. J. A]
  {10.1140/epja/s10050-019-00018-6}, 56, 15

\bibitem[\protect\citeauthoryear{Fattoyev, Horowitz, Piekarewicz  \&
  Shen}{Fattoyev et~al.}{2010}]{Fattoyev:2010mx}
Fattoyev F.~J.,  Horowitz C.~J.,  Piekarewicz J.,   Shen G.,  2010, \mn@doi
  [Phys. Rev. C] {10.1103/PhysRevC.82.055803}, 82, 055803

\bibitem[\protect\citeauthoryear{{Finzi} \& {Wolf}}{{Finzi} \&
  {Wolf}}{1968}]{1968ApJ...153..835F}
{Finzi} A.,  {Wolf} R.~A.,  1968, \mn@doi [\apj] {10.1086/149708}, \href
  {https://ui.adsabs.harvard.edu/abs/1968ApJ...153..835F} {153, 835}

\bibitem[\protect\citeauthoryear{Flanagan \& Hinderer}{Flanagan \&
  Hinderer}{2008}]{Flanagan:2007ix}
Flanagan E.~E.,  Hinderer T.,  2008, \mn@doi [Phys. Rev. D]
  {10.1103/PhysRevD.77.021502}, 77, 021502

\bibitem[\protect\citeauthoryear{Florkowski \& Ryblewski}{Florkowski \&
  Ryblewski}{2011}]{Florkowski:2010cf}
Florkowski W.,  Ryblewski R.,  2011, \mn@doi [Phys. Rev. C]
  {10.1103/PhysRevC.83.034907}, 83, 034907

\bibitem[\protect\citeauthoryear{Gavassino, Antonelli  \& Haskell}{Gavassino
  et~al.}{2021}]{Gavassino:2020kwo}
Gavassino L.,  Antonelli M.,   Haskell B.,  2021, \mn@doi [Class. Quant. Grav.]
  {10.1088/1361-6382/abe588}, 38, 075001

\bibitem[\protect\citeauthoryear{Giacalone, Noronha-Hostler, Luzum  \&
  Ollitrault}{Giacalone et~al.}{2018}]{Giacalone:2017dud}
Giacalone G.,  Noronha-Hostler J.,  Luzum M.,   Ollitrault J.-Y.,  2018,
  \mn@doi [Phys. Rev. C] {10.1103/PhysRevC.97.034904}, 97, 034904

\bibitem[\protect\citeauthoryear{Glendenning}{Glendenning}{1997}]{Glendenning:1997wn}
Glendenning N.~K.,  1997, {Compact stars: Nuclear physics, particle physics,
  and general relativity}

\bibitem[\protect\citeauthoryear{Guillot, Pavlov, Reyes, Reisenegger,
  Rodriguez, Rangelov  \& Kargaltsev}{Guillot et~al.}{2019}]{Guillot:2019ugf}
Guillot S.,  Pavlov G.~G.,  Reyes C.,  Reisenegger A.,  Rodriguez L.,  Rangelov
  B.,   Kargaltsev O.,  2019, \mn@doi [Astrophys. J.]
  {10.3847/1538-4357/ab0f38}, 874, 175

\bibitem[\protect\citeauthoryear{Gupta, Singh, Anand  \& Wadhwa}{Gupta
  et~al.}{1997}]{Gupta:1997ce}
Gupta V.~K.,  Singh S.,  Anand J.~D.,   Wadhwa A.,  1997, \mn@doi [Pramana]
  {10.1007/BF02847431}, 49, 443

\bibitem[\protect\citeauthoryear{Haensel \& Schaeffer}{Haensel \&
  Schaeffer}{1992}]{Haensel:1992zz}
Haensel P.,  Schaeffer R.,  1992, \mn@doi [Phys. Rev. D]
  {10.1103/PhysRevD.45.4708}, 45, 4708

\bibitem[\protect\citeauthoryear{Haensel, Levenfish  \& Yakovlev}{Haensel
  et~al.}{2000}]{Haensel:2000vz}
Haensel P.,  Levenfish K.~P.,   Yakovlev D.~G.,  2000, Astron. Astrophys., 357,
  1157

\bibitem[\protect\citeauthoryear{Haensel, Levenfish  \& Yakovlev}{Haensel
  et~al.}{2001}]{Haensel:2001mw}
Haensel P.,  Levenfish K.~P.,   Yakovlev D.~G.,  2001, \mn@doi [Astron.
  Astrophys.] {10.1051/0004-6361:20010383}, 327, 130

\bibitem[\protect\citeauthoryear{Hanauske, Takami, Bovard, Rezzolla, Font,
  Galeazzi  \& St\"ocker}{Hanauske et~al.}{2017}]{Hanauske:2016gia}
Hanauske M.,  Takami K.,  Bovard L.,  Rezzolla L.,  Font J.~A.,  Galeazzi F.,
  St\"ocker H.,  2017, \mn@doi [Phys. Rev. D] {10.1103/PhysRevD.96.043004}, 96,
  043004

\bibitem[\protect\citeauthoryear{Harutyunyan \& Sedrakian}{Harutyunyan \&
  Sedrakian}{2017}]{Harutyunyan:2017ieu}
Harutyunyan A.,  Sedrakian A.,  2017, \mn@doi [Phys. Rev. D]
  {10.1103/PhysRevD.96.034006}, 96, 034006

\bibitem[\protect\citeauthoryear{Hoult \& Kovtun}{Hoult \&
  Kovtun}{2020}]{Hoult:2020eho}
Hoult R.~E.,  Kovtun P.,  2020, \mn@doi [JHEP] {10.1007/JHEP06(2020)067}, 06,
  067

\bibitem[\protect\citeauthoryear{Israel \& Stewart}{Israel \&
  Stewart}{1979}]{Israel:1979wp}
Israel W.,  Stewart J.~M.,  1979, \mn@doi [Annals Phys.]
  {10.1016/0003-4916(79)90130-1}, 118, 341

\bibitem[\protect\citeauthoryear{Kastaun, Ciolfi  \& Giacomazzo}{Kastaun
  et~al.}{2016}]{Kastaun:2016yaf}
Kastaun W.,  Ciolfi R.,   Giacomazzo B.,  2016, \mn@doi [Phys. Rev. D]
  {10.1103/PhysRevD.94.044060}, 94, 044060

\bibitem[\protect\citeauthoryear{Kokkotas \& Schmidt}{Kokkotas \&
  Schmidt}{1999}]{Kokkotas:1999bd}
Kokkotas K.~D.,  Schmidt B.~G.,  1999, \mn@doi [Living Rev. Rel.]
  {10.12942/lrr-1999-2}, 2, 2

\bibitem[\protect\citeauthoryear{Kolomeitsev \& Voskresensky}{Kolomeitsev \&
  Voskresensky}{2015}]{Kolomeitsev:2014gfa}
Kolomeitsev E.~E.,  Voskresensky D.~N.,  2015, \mn@doi [Phys. Rev. C]
  {10.1103/PhysRevC.91.025805}, 91, 025805

\bibitem[\protect\citeauthoryear{Kovtun}{Kovtun}{2019}]{Kovtun:2019hdm}
Kovtun P.,  2019, \mn@doi [JHEP] {10.1007/JHEP10(2019)034}, 10, 034

\bibitem[\protect\citeauthoryear{Lai}{Lai}{1994}]{Lai:1993di}
Lai D.,  1994, \mn@doi [Mon. Not. Roy. Astron. Soc.] {10.1093/mnras/270.3.611},
  270, 611

\bibitem[\protect\citeauthoryear{Lai}{Lai}{2001}]{Lai:2001jt}
Lai D.,  2001, \mn@doi [AIP Conf. Proc.] {10.1063/1.1387316}, 575, 246

\bibitem[\protect\citeauthoryear{Lattimer \& Prakash}{Lattimer \&
  Prakash}{2016}]{Lattimer:2015nhk}
Lattimer J.~M.,  Prakash M.,  2016, \mn@doi [Phys. Rept.]
  {10.1016/j.physrep.2015.12.005}, 621, 127

\bibitem[\protect\citeauthoryear{Lin, Caplan, Horowitz  \& Lunardini}{Lin
  et~al.}{2020}]{Lin:2020nxy}
Lin Z.,  Caplan M.~E.,  Horowitz C.~J.,   Lunardini C.,  2020, \mn@doi [Phys.
  Rev. C] {10.1103/PhysRevC.102.045801}, 102, 045801

\bibitem[\protect\citeauthoryear{Liu, Greco, Baran, Colonna  \& Di~Toro}{Liu
  et~al.}{2002}]{Liu:2001iz}
Liu B.,  Greco V.,  Baran V.,  Colonna M.,   Di~Toro M.,  2002, \mn@doi [Phys.
  Rev. C] {10.1103/PhysRevC.65.045201}, 65, 045201

\bibitem[\protect\citeauthoryear{Madsen}{Madsen}{1992}]{Madsen:1992sx}
Madsen J.,  1992, \mn@doi [Phys. Rev.] {10.1103/PhysRevD.46.3290}, D46, 3290

\bibitem[\protect\citeauthoryear{Margalit \& Metzger}{Margalit \&
  Metzger}{2017}]{Margalit:2017dij}
Margalit B.,  Metzger B.~D.,  2017, \mn@doi [Astrophys. J. Lett.]
  {10.3847/2041-8213/aa991c}, 850, L19

\bibitem[\protect\citeauthoryear{Martinez \& Strickland}{Martinez \&
  Strickland}{2010}]{Martinez:2010sc}
Martinez M.,  Strickland M.,  2010, \mn@doi [Nucl. Phys. A]
  {10.1016/j.nuclphysa.2010.08.011}, 848, 183

\bibitem[\protect\citeauthoryear{Metzger}{Metzger}{2020}]{Metzger:2019zeh}
Metzger B.~D.,  2020, \mn@doi [Living Rev. Rel.] {10.1007/s41114-019-0024-0},
  23, 1

\bibitem[\protect\citeauthoryear{Miller et~al.}{Miller
  et~al.}{2019}]{Miller:2019cac}
Miller M.~C.,  et~al., 2019, \mn@doi [Astrophys. J. Lett.]
  {10.3847/2041-8213/ab50c5}, 887, L24

\bibitem[\protect\citeauthoryear{Miller et~al.}{Miller
  et~al.}{2021}]{Miller:2021qha}
Miller M.~C.,  et~al., 2021

\bibitem[\protect\citeauthoryear{Misner, Thorne  \& Wheeler}{Misner
  et~al.}{1973}]{Misner:1973prb}
Misner C.~W.,  Thorne K.~S.,   Wheeler J.~A.,  1973, {Gravitation}.
W. H. Freeman, San Francisco

\bibitem[\protect\citeauthoryear{Monnai \& Hirano}{Monnai \&
  Hirano}{2009}]{Monnai:2009ad}
Monnai A.,  Hirano T.,  2009, \mn@doi [Phys. Rev. C]
  {10.1103/PhysRevC.80.054906}, 80, 054906

\bibitem[\protect\citeauthoryear{Moreland, Bernhard  \& Bass}{Moreland
  et~al.}{2015}]{Moreland:2014oya}
Moreland J.~S.,  Bernhard J.~E.,   Bass S.~A.,  2015, \mn@doi [Phys. Rev. C]
  {10.1103/PhysRevC.92.011901}, 92, 011901

\bibitem[\protect\citeauthoryear{Most, Weih, Rezzolla  \&
  Schaffner-Bielich}{Most et~al.}{2018}]{Most:2018hfd}
Most E.~R.,  Weih L.~R.,  Rezzolla L.,   Schaffner-Bielich J.,  2018, \mn@doi
  [Phys. Rev. Lett.] {10.1103/PhysRevLett.120.261103}, 120, 261103

\bibitem[\protect\citeauthoryear{Most, Papenfort, Dexheimer, Hanauske, Schramm,
  St\"ocker  \& Rezzolla}{Most et~al.}{2019a}]{Most:2018eaw}
Most E.~R.,  Papenfort L.~J.,  Dexheimer V.,  Hanauske M.,  Schramm S.,
  St\"ocker H.,   Rezzolla L.,  2019a, \mn@doi [Phys. Rev. Lett.]
  {10.1103/PhysRevLett.122.061101}, 122, 061101

\bibitem[\protect\citeauthoryear{Most, Papenfort  \& Rezzolla}{Most
  et~al.}{2019b}]{Most:2019kfe}
Most E.~R.,  Papenfort L.~J.,   Rezzolla L.,  2019b, \mn@doi [Mon. Not. Roy.
  Astron. Soc.] {10.1093/mnras/stz2809}, 490, 3588

\bibitem[\protect\citeauthoryear{Most, Jens~Papenfort, Dexheimer, Hanauske,
  Stoecker  \& Rezzolla}{Most et~al.}{2020a}]{Most:2019onn}
Most E.~R.,  Jens~Papenfort L.,  Dexheimer V.,  Hanauske M.,  Stoecker H.,
  Rezzolla L.,  2020a, \mn@doi [Eur. Phys. J. A]
  {10.1140/epja/s10050-020-00073-4}, 56, 59

\bibitem[\protect\citeauthoryear{Most, Papenfort, Weih  \& Rezzolla}{Most
  et~al.}{2020b}]{Most:2020bba}
Most E.~R.,  Papenfort L.~J.,  Weih L.~R.,   Rezzolla L.,  2020b, \mn@doi [Mon.
  Not. Roy. Astron. Soc.] {10.1093/mnrasl/slaa168}, 499, L82

\bibitem[\protect\citeauthoryear{Motornenko, Steinheimer  \&
  Stoecker}{Motornenko et~al.}{2021}]{Motornenko:2021dbq}
Motornenko A.,  Steinheimer J.,   Stoecker H.,  2021.  (\mn@eprint {arXiv}
  {2105.12475})

\bibitem[\protect\citeauthoryear{Mueller, Janka  \& Marek}{Mueller
  et~al.}{2013}]{Mueller:2012sv}
Mueller B.,  Janka H.-T.,   Marek A.,  2013, \mn@doi [Astrophys. J.]
  {10.1088/0004-637X/766/1/43}, 766, 43

\bibitem[\protect\citeauthoryear{Nathanail, Most  \& Rezzolla}{Nathanail
  et~al.}{2021}]{Nathanail:2021tay}
Nathanail A.,  Most E.~R.,   Rezzolla L.,  2021, \mn@doi [Astrophys. J. Lett.]
  {10.3847/2041-8213/abdfc6}, 908, L28

\bibitem[\protect\citeauthoryear{N\"attil\"a, Miller, Steiner, Kajava,
  Suleimanov  \& Poutanen}{N\"attil\"a et~al.}{2017}]{Nattila:2017wtj}
N\"attil\"a J.,  Miller M.~C.,  Steiner A.~W.,  Kajava J. J.~E.,  Suleimanov
  V.~F.,   Poutanen J.,  2017, \mn@doi [Astron. Astrophys.]
  {10.1051/0004-6361/201731082}, 608, A31

\bibitem[\protect\citeauthoryear{Niemi, Eskola, Paatelainen  \& Tuominen}{Niemi
  et~al.}{2016a}]{Niemi:2015voa}
Niemi H.,  Eskola K.~J.,  Paatelainen R.,   Tuominen K.,  2016a, \mn@doi [Phys.
  Rev. C] {10.1103/PhysRevC.93.014912}, 93, 014912

\bibitem[\protect\citeauthoryear{Niemi, Eskola  \& Paatelainen}{Niemi
  et~al.}{2016b}]{Niemi:2015qia}
Niemi H.,  Eskola K.~J.,   Paatelainen R.,  2016b, \mn@doi [Phys. Rev. C]
  {10.1103/PhysRevC.93.024907}, 93, 024907

\bibitem[\protect\citeauthoryear{Noronha-Hostler, Denicol, Noronha, Andrade  \&
  Grassi}{Noronha-Hostler et~al.}{2013}]{Noronha-Hostler:2013gga}
Noronha-Hostler J.,  Denicol G.~S.,  Noronha J.,  Andrade R. P.~G.,   Grassi
  F.,  2013, \mn@doi [Phys. Rev. C] {10.1103/PhysRevC.88.044916}, 88, 044916

\bibitem[\protect\citeauthoryear{Noronha-Hostler, Noronha  \&
  Grassi}{Noronha-Hostler et~al.}{2014}]{Noronha-Hostler:2014dqa}
Noronha-Hostler J.,  Noronha J.,   Grassi F.,  2014, \mn@doi [Phys. Rev. C]
  {10.1103/PhysRevC.90.034907}, 90, 034907

\bibitem[\protect\citeauthoryear{Noronha-Hostler, Yan, Gardim  \&
  Ollitrault}{Noronha-Hostler et~al.}{2016}]{Noronha-Hostler:2015dbi}
Noronha-Hostler J.,  Yan L.,  Gardim F.~G.,   Ollitrault J.-Y.,  2016, \mn@doi
  [Phys. Rev. C] {10.1103/PhysRevC.93.014909}, 93, 014909

\bibitem[\protect\citeauthoryear{\"Ozel \& Freire}{\"Ozel \&
  Freire}{2016}]{Ozel:2016oaf}
\"Ozel F.,  Freire P.,  2016, \mn@doi [Ann. Rev. Astron. Astrophys.]
  {10.1146/annurev-astro-081915-023322}, 54, 401

\bibitem[\protect\citeauthoryear{{{\"O}zel} \& {Psaltis}}{{{\"O}zel} \&
  {Psaltis}}{2009}]{2009PhRvD..80j3003O}
{{\"O}zel} F.,  {Psaltis} D.,  2009, \mn@doi [\prd]
  {10.1103/PhysRevD.80.103003}, \href
  {https://ui.adsabs.harvard.edu/abs/2009PhRvD..80j3003O} {80, 103003}

\bibitem[\protect\citeauthoryear{Ozel, Baym  \& Guver}{Ozel
  et~al.}{2010}]{Ozel:2010fw}
Ozel F.,  Baym G.,   Guver T.,  2010, \mn@doi [Phys. Rev. D]
  {10.1103/PhysRevD.82.101301}, 82, 101301

\bibitem[\protect\citeauthoryear{Page \& Reddy}{Page \&
  Reddy}{2006}]{Page:2006ud}
Page D.,  Reddy S.,  2006, \mn@doi [Ann. Rev. Nucl. Part. Sci.]
  {10.1146/annurev.nucl.56.080805.140600}, 56, 327

\bibitem[\protect\citeauthoryear{Pandya \& Pretorius}{Pandya \&
  Pretorius}{2021}]{Pandya:2021ief}
Pandya A.,  Pretorius F.,  2021

\bibitem[\protect\citeauthoryear{Perego, Bernuzzi  \& Radice}{Perego
  et~al.}{2019}]{Perego:2019adq}
Perego A.,  Bernuzzi S.,   Radice D.,  2019, \mn@doi [Eur. Phys. J. A]
  {10.1140/epja/i2019-12810-7}, 55, 124

\bibitem[\protect\citeauthoryear{Peters}{Peters}{1964}]{Peters:1964zz}
Peters P.~C.,  1964, \mn@doi [Phys. Rev.] {10.1103/PhysRev.136.B1224}, 136,
  B1224

\bibitem[\protect\citeauthoryear{Petersen, Steinheimer, Burau, Bleicher  \&
  St\"ocker}{Petersen et~al.}{2008}]{Petersen:2008dd}
Petersen H.,  Steinheimer J.,  Burau G.,  Bleicher M.,   St\"ocker H.,  2008,
  \mn@doi [Phys. Rev. C] {10.1103/PhysRevC.78.044901}, 78, 044901

\bibitem[\protect\citeauthoryear{Philipsen}{Philipsen}{2013}]{Philipsen:2012nu}
Philipsen O.,  2013, \mn@doi [Prog. Part. Nucl. Phys.]
  {10.1016/j.ppnp.2012.09.003}, 70, 55

\bibitem[\protect\citeauthoryear{Plumberg, Almaalol, Dore, Noronha  \&
  Noronha-Hostler}{Plumberg et~al.}{2021}]{Plumberg:2021bme}
Plumberg C.,  Almaalol D.,  Dore T.,  Noronha J.,   Noronha-Hostler J.,  2021

\bibitem[\protect\citeauthoryear{Radice}{Radice}{2017}]{Radice:2017zta}
Radice D.,  2017, \mn@doi [Astrophys. J. Lett.] {10.3847/2041-8213/aa6483},
  838, L2

\bibitem[\protect\citeauthoryear{Radice, Bernuzzi, Del~Pozzo, Roberts  \&
  Ott}{Radice et~al.}{2017}]{Radice:2016rys}
Radice D.,  Bernuzzi S.,  Del~Pozzo W.,  Roberts L.~F.,   Ott C.~D.,  2017,
  \mn@doi [Astrophys. J. Lett.] {10.3847/2041-8213/aa775f}, 842, L10

\bibitem[\protect\citeauthoryear{Raithel, \"Ozel  \& Psaltis}{Raithel
  et~al.}{2018}]{Raithel:2018ncd}
Raithel C.,  \"Ozel F.,   Psaltis D.,  2018, \mn@doi [Astrophys. J. Lett.]
  {10.3847/2041-8213/aabcbf}, 857, L23

\bibitem[\protect\citeauthoryear{Raithel, Paschalidis  \& \"Ozel}{Raithel
  et~al.}{2021}]{Raithel:2021hye}
Raithel C.~A.,  Paschalidis V.,   \"Ozel F.,  2021

\bibitem[\protect\citeauthoryear{Rajagopal \& Tripuraneni}{Rajagopal \&
  Tripuraneni}{2010}]{Rajagopal:2009yw}
Rajagopal K.,  Tripuraneni N.,  2010, \mn@doi [JHEP] {10.1007/JHEP03(2010)018},
  03, 018

\bibitem[\protect\citeauthoryear{Read, Markakis, Shibata, Uryu, Creighton  \&
  Friedman}{Read et~al.}{2009}]{Read:2009yp}
Read J.~S.,  Markakis C.,  Shibata M.,  Uryu K.,  Creighton J. D.~E.,
  Friedman J.~L.,  2009, \mn@doi [Phys. Rev. D] {10.1103/PhysRevD.79.124033},
  79, 124033

\bibitem[\protect\citeauthoryear{Reed, Fattoyev, Horowitz  \& Piekarewicz}{Reed
  et~al.}{2021}]{Reed:2021nqk}
Reed B.~T.,  Fattoyev F.~J.,  Horowitz C.~J.,   Piekarewicz J.,  2021, \mn@doi
  [Phys. Rev. Lett.] {10.1103/PhysRevLett.126.172503}, 126, 172503

\bibitem[\protect\citeauthoryear{Reisenegger}{Reisenegger}{1995}]{Reisenegger:1994be}
Reisenegger A.,  1995, \mn@doi [Astrophys. J.] {10.1086/175480}, 442, 749

\bibitem[\protect\citeauthoryear{Rezzolla \& Zanotti}{Rezzolla \&
  Zanotti}{2013}]{Rezzolla_Zanotti_book}
Rezzolla L.,  Zanotti O.,  2013, Relativistic hydrodynamics.
Oxford University Press, New York

\bibitem[\protect\citeauthoryear{Rezzolla, Most  \& Weih}{Rezzolla
  et~al.}{2018}]{Rezzolla:2017aly}
Rezzolla L.,  Most E.~R.,   Weih L.~R.,  2018, \mn@doi [Astrophys. J. Lett.]
  {10.3847/2041-8213/aaa401}, 852, L25

\bibitem[\protect\citeauthoryear{Riley et~al.}{Riley
  et~al.}{2019}]{Riley:2019yda}
Riley T.~E.,  et~al., 2019, \mn@doi [Astrophys. J. Lett.]
  {10.3847/2041-8213/ab481c}, 887, L21

\bibitem[\protect\citeauthoryear{Riley et~al.}{Riley
  et~al.}{2021}]{Riley:2021pdl}
Riley T.~E.,  et~al., 2021

\bibitem[\protect\citeauthoryear{Roberts \& Reddy}{Roberts \&
  Reddy}{2017}]{Roberts:2016mwj}
Roberts L.~F.,  Reddy S.,  2017, \mn@doi [Phys. Rev. C]
  {10.1103/PhysRevC.95.045807}, 95, 045807

\bibitem[\protect\citeauthoryear{Romatschke \& Romatschke}{Romatschke \&
  Romatschke}{2019}]{Romatschke:2017ejr}
Romatschke P.,  Romatschke U.,  2019, {Relativistic Fluid Dynamics In and Out
  of Equilibrium}.
Cambridge Monographs on Mathematical Physics, Cambridge University Press
  (\mn@eprint {arXiv} {1712.05815}), \mn@doi{10.1017/9781108651998}

\bibitem[\protect\citeauthoryear{Ruiz, Shapiro  \& Tsokaros}{Ruiz
  et~al.}{2018}]{Ruiz:2017due}
Ruiz M.,  Shapiro S.~L.,   Tsokaros A.,  2018, \mn@doi [Phys. Rev. D]
  {10.1103/PhysRevD.97.021501}, 97, 021501

\bibitem[\protect\citeauthoryear{Ryu, Paquet, Shen, Denicol, Schenke, Jeon  \&
  Gale}{Ryu et~al.}{2015}]{Ryu:2015vwa}
Ryu S.,  Paquet J.~F.,  Shen C.,  Denicol G.~S.,  Schenke B.,  Jeon S.,   Gale
  C.,  2015, \mn@doi [Phys. Rev. Lett.] {10.1103/PhysRevLett.115.132301}, 115,
  132301

\bibitem[\protect\citeauthoryear{Ryu, Paquet, Shen, Denicol, Schenke, Jeon  \&
  Gale}{Ryu et~al.}{2018}]{Ryu:2017qzn}
Ryu S.,  Paquet J.-F.,  Shen C.,  Denicol G.,  Schenke B.,  Jeon S.,   Gale C.,
   2018, \mn@doi [Phys. Rev. C] {10.1103/PhysRevC.97.034910}, 97, 034910

\bibitem[\protect\citeauthoryear{Sawyer}{Sawyer}{1980}]{Sawyer:1980wp}
Sawyer R.~F.,  1980, \mn@doi [Astrophys. J.] {10.1086/157858}, 237, 187

\bibitem[\protect\citeauthoryear{Sawyer}{Sawyer}{1989}]{Sawyer:1989dp}
Sawyer R.~F.,  1989, \mn@doi [Phys. Rev. D] {10.1103/PhysRevD.39.3804}, 39,
  3804

\bibitem[\protect\citeauthoryear{Schenke}{Schenke}{2021}]{Schenke:2021mxx}
Schenke B.,  2021

\bibitem[\protect\citeauthoryear{Schenke, Shen  \& Tribedy}{Schenke
  et~al.}{2020}]{Schenke:2020mbo}
Schenke B.,  Shen C.,   Tribedy P.,  2020, \mn@doi [Phys. Rev. C]
  {10.1103/PhysRevC.102.044905}, 102, 044905

\bibitem[\protect\citeauthoryear{Schmitt \& Shternin}{Schmitt \&
  Shternin}{2018}]{Schmitt:2017efp}
Schmitt A.,  Shternin P.,  2018, \mn@doi [Astrophys. Space Sci. Libr.]
  {10.1007/978-3-319-97616-7_9}, 457, 455

\bibitem[\protect\citeauthoryear{Sekiguchi, Kiuchi, Kyutoku  \&
  Shibata}{Sekiguchi et~al.}{2011}]{Sekiguchi:2011mc}
Sekiguchi Y.,  Kiuchi K.,  Kyutoku K.,   Shibata M.,  2011, \mn@doi [Phys. Rev.
  Lett.] {10.1103/PhysRevLett.107.211101}, 107, 211101

\bibitem[\protect\citeauthoryear{Shen \& Yan}{Shen \& Yan}{2020}]{Shen:2020mgh}
Shen C.,  Yan L.,  2020, \mn@doi [Nucl. Sci. Tech.]
  {10.1007/s41365-020-00829-z}, 31, 122

\bibitem[\protect\citeauthoryear{Shen, Qiu, Song, Bernhard, Bass  \&
  Heinz}{Shen et~al.}{2016}]{Shen:2014vra}
Shen C.,  Qiu Z.,  Song H.,  Bernhard J.,  Bass S.,   Heinz U.,  2016, \mn@doi
  [Comput. Phys. Commun.] {10.1016/j.cpc.2015.08.039}, 199, 61

\bibitem[\protect\citeauthoryear{Shibata \& Kiuchi}{Shibata \&
  Kiuchi}{2017}]{Shibata:2017xht}
Shibata M.,  Kiuchi K.,  2017, \mn@doi [Phys. Rev. D]
  {10.1103/PhysRevD.95.123003}, 95, 123003

\bibitem[\protect\citeauthoryear{Shibata, Zhou, Kiuchi  \& Fujibayashi}{Shibata
  et~al.}{2019}]{Shibata:2019ctb}
Shibata M.,  Zhou E.,  Kiuchi K.,   Fujibayashi S.,  2019, \mn@doi [Phys. Rev.
  D] {10.1103/PhysRevD.100.023015}, 100, 023015

\bibitem[\protect\citeauthoryear{Sirunyan et~al.}{Sirunyan
  et~al.}{2019}]{Sirunyan:2019wqp}
Sirunyan A.~M.,  et~al., 2019, \mn@doi [Phys. Rev. C]
  {10.1103/PhysRevC.100.044902}, 100, 044902

\bibitem[\protect\citeauthoryear{Song \& Heinz}{Song \&
  Heinz}{2010}]{Song:2009rh}
Song H.,  Heinz U.~W.,  2010, \mn@doi [Phys. Rev. C]
  {10.1103/PhysRevC.81.024905}, 81, 024905

\bibitem[\protect\citeauthoryear{Steiner, Lattimer  \& Brown}{Steiner
  et~al.}{2010}]{Steiner:2010fz}
Steiner A.~W.,  Lattimer J.~M.,   Brown E.~F.,  2010, \mn@doi [Astrophys. J.]
  {10.1088/0004-637X/722/1/33}, 722, 33

\bibitem[\protect\citeauthoryear{Summerfield, Lu, Plumberg, Lee,
  Noronha-Hostler  \& Timmins}{Summerfield et~al.}{2021}]{Summerfield:2021oex}
Summerfield N.,  Lu B.-N.,  Plumberg C.,  Lee D.,  Noronha-Hostler J.,
  Timmins A.,  2021

\bibitem[\protect\citeauthoryear{Sykes \& Brooker}{Sykes \&
  Brooker}{1970}]{SYKES19701}
Sykes J.,  Brooker G.,  1970, \mn@doi [Annals of Physics]
  {https://doi.org/10.1016/0003-4916(70)90002-3}, 56, 1

\bibitem[\protect\citeauthoryear{Takami, Rezzolla  \& Baiotti}{Takami
  et~al.}{2014}]{Takami:2014zpa}
Takami K.,  Rezzolla L.,   Baiotti L.,  2014, \mn@doi [Phys. Rev. Lett.]
  {10.1103/PhysRevLett.113.091104}, 113, 091104

\bibitem[\protect\citeauthoryear{Takami, Rezzolla  \& Baiotti}{Takami
  et~al.}{2015}]{Takami:2014tva}
Takami K.,  Rezzolla L.,   Baiotti L.,  2015, \mn@doi [Phys. Rev. D]
  {10.1103/PhysRevD.91.064001}, 91, 064001

\bibitem[\protect\citeauthoryear{Torrieri \& Mishustin}{Torrieri \&
  Mishustin}{2008}]{Torrieri:2008ip}
Torrieri G.,  Mishustin I.,  2008, \mn@doi [Phys. Rev. C]
  {10.1103/PhysRevC.78.021901}, 78, 021901

\bibitem[\protect\citeauthoryear{Weller \& Romatschke}{Weller \&
  Romatschke}{2017}]{Weller:2017tsr}
Weller R.~D.,  Romatschke P.,  2017, \mn@doi [Phys. Lett. B]
  {10.1016/j.physletb.2017.09.077}, 774, 351

\bibitem[\protect\citeauthoryear{Yakovlev, Kaminker, Gnedin  \&
  Haensel}{Yakovlev et~al.}{2001}]{Yakovlev:2000jp}
Yakovlev D.~G.,  Kaminker A.~D.,  Gnedin O.~Y.,   Haensel P.,  2001, \mn@doi
  [Phys. Rept.] {10.1016/S0370-1573(00)00131-9}, 354, 1

\bibitem[\protect\citeauthoryear{Yakovlev, Gusakov  \& Haensel}{Yakovlev
  et~al.}{2018}]{Yakovlev:2018jia}
Yakovlev D.~G.,  Gusakov M.~E.,   Haensel P.,  2018, \mn@doi [Mon. Not. Roy.
  Astron. Soc.] {10.1093/mnras/sty2639}, 481, 4924

\makeatother
\end{thebibliography}
%
%
%
%
%
\appendix
%
%
\bsp	
\label{lastpage}
\end{document}